\documentclass[journal,10pt]{IEEEtran}

\usepackage{amssymb}
\usepackage{amsmath}

\usepackage{url}
\usepackage{xcolor}
\usepackage{cite,graphicx,amsmath,amssymb}
\usepackage{subfigure}
\usepackage{citesort}
\usepackage{fancyhdr}
\usepackage{mdwmath}
\usepackage{mdwtab}
\usepackage{caption}
\usepackage{amsthm}
\usepackage{amsmath}
\usepackage{bbm}
\usepackage{epstopdf}
\usepackage{cite}
\usepackage{multirow}

\newtheorem{remark}{Remark}
\newtheorem{theorem}{Theorem}

\newtheorem{lemma}{Lemma}

\newtheorem{corollary}{Corollary}

\makeatletter
\def\ScaleIfNeeded{%
\ifdim\Gin@nat@width>\linewidth \linewidth \else \Gin@nat@width
\fi } \makeatother

\begin{document}

\title{Semi-Integrated-Sensing-and-Communication (Semi-ISaC): From OMA to NOMA}

\author{Chao~Zhang,~\IEEEmembership{Graduate Student Member,~IEEE,}
        Wenqiang~Yi,~\IEEEmembership{Member,~IEEE,}
        Yuanwei~Liu,~\IEEEmembership{Senior Member,~IEEE,}
        and Lajos~Hanzo,~\IEEEmembership{Life Fellow,~IEEE}

\thanks{Lajos Hanzo would like to acknowledge the financial support of the Engineering and Physical Sciences Research Council projects EP/W016605/1 and EP/X01228X/1 as well as of the European Research Council's Advanced Fellow Grant QuantCom (Grant No. 789028).}
\thanks{Chao Zhang, Wenqiang Yi, and Yuanwei Liu are the School of Electronic Engineering and Computer Science, Queen Mary University of London, London E1 4NS, U.K. (email:\{chao.zhang, w.yi, yuanwei.liu\}@qmul.ac.uk).}
\thanks{Lajos Hanzo is with the School of Electronics and Computer Science, University of Southampton, Southampton, SO17 1BJ, U.K. (e-mail: lh@ecs.soton.ac.uk).}
\thanks{Part of this work has been accepted to IEEE International Conference on Communications (ICC), Seoul, South Korea, May, 2022 \cite{conf}.}

}

\maketitle

\begin{abstract}
  The new concept of semi-integrated-sensing-and-communication (Semi-ISaC) is proposed for next-generation cellular networks. Compared to the state-of-the-art, where the total bandwidth is used for integrated sensing and communication (ISaC), the proposed Semi-ISaC framework provides more freedom as it allows that a portion of the bandwidth is exclusively used for either wireless communication or radar detection, while the rest is for ISaC transmission. To enhance the bandwidth efficiency (BE), we investigate the evolution of Semi-ISaC networks from orthogonal multiple access (OMA) to non-orthogonal multiple access (NOMA). First, we evaluate the performance of an OMA-based Semi-ISaC network. As for the communication signals, we investigate both the outage probability (OP) and the ergodic rate. As for the radar echoes, we characterize the ergodic radar estimation information rate (REIR). Then, we investigate the performance of a NOMA-based Semi-ISaC network, including the OP and the ergodic rate for communication signals and the ergodic REIR for radar echoes. The diversity gains of OP and the high signal-to-noise ratio (SNR) slopes of the ergodic REIR are also evaluated as insights. The analytical results indicate that: 1) Under a two-user NOMA-based Semi-ISaC scenario, the diversity order of the near-user is equal to the coefficient of the Nakagami-\emph{m} fading channels ($m$), while that of the far-user is zero; and 2) The high-SNR slope for the ergodic REIR is based on the ratio of the radar signal's duty cycle to the pulse duration. Our simulation results show that: 1) Semi-ISaC has better channel capacity than the conventional ISaC; and 2) The NOMA-based Semi-ISaC has better channel capacity than the OMA-based Semi-ISaC.
\end{abstract}


\section{Introduction}

Given the incessant escalation of wireless tele-traffic, the impending spectrum-crunch can only be circumvented by the migration to millimeter-wave (mm-wave) carriers. However, since radar sensing technologies also rely on mm-wave carriers, the bandwidth of sensing and wireless communication might become overlapped \cite{ISAC_basic2,ISAC_basic,ISACJ}. Indeed, it is possible to economize by sophisticated bandwidth-sharing in next-generation wireless communications (6G) with the aid of integrated sensing and communication (ISaC) \cite{servy}.

In practical scenarios, it is difficult to exploit the total spectrum for ISaC as the bandwidth has already been occupied by different applications, as exemplified by the L-band (1-2 GHz) for long-range air traffic control and long-range surveillance; the S-band (2-4 GHz) for terminal air traffic control, moderate-range surveillance, and long-range weather observation; the C-band (4-8 GHz) for long-range tracking, weapon location, and weather observation; and the mm-waves for high-resolution mapping, satellite altimetry, vehicular radars, and police radars \cite{band}. Hence, the most practical scenario is that a given portion of the bandwidth is exploited for ISaC in support of its specific applications, while the remaining bandwidth is exploited only for bandwidth-thirsty wireless communications or radar detection. Explicitly, compared to conventional ISaC scenarios, a semi-integrated-sensing-and-communication (Semi-ISaC) solution is more promising for next-generation applications.

As for Semi-ISaC, the communication and sensing signals are superimposed since the communication and the radar systems share the same resource blocks. Sensing needs continuous waves that spans a large range of bandwidth. Hence, due to the shortage of spectral resources, we need multiple access schemes, such as non-orthogonal multiple access (NOMA). We are also able to exploit NOMA to improve the bandwidth efficiency (BE) compared to orthogonal multiple access (OMA). Additionally, since the propagation properties of sensing and wireless communication are different (even for the same user), we always have a gap between the power levels of the sensing and communication functions at the receiver side. As for the case associated with significant power differences, the power domain NOMA can perform well in terms of supporting multiple users or functions. Moreover, the successive interference cancellation (SIC) scheme has been exploited in uplink ISaC networks to mitigate the interference of the overlapped signals \cite{yuanweiNOMA,ISACjournal}. By harnessing SIC, the BE of ISaC systems significantly increases compared to the cases associated with full radar echo interference in the overlapped period \cite{ISACjournal}. Furthermore, NOMA provides a new degree of freedom for the Semi-ISaC networks to obtain high flexibility for further designs. To sum up, the power domain NOMA is an eminently suitable solution for multiple access in the Semi-ISaC networks.

\subsection{Related Works}

Based on the advantages of NOMA in Semi-ISaC networks, the NOMA-based Semi-ISaC networks have attracted increasing attention both in academia and in industry. The past decades have witnessed the development of NOMA \cite{yuanweiNOMA} and more recently of ISaC \cite{ISaC_Magzine,ISACjournal}, but there is a paucity of literature on their amalgam.

\subsubsection{Related literature of NOMA}

Several decades have passed since the concept of NOMA was first proposed. But in the recent five years, power domain NOMA combined with SIC and power allocation methods has gained popularity \cite{yuanwei2,zhiguo,SIC2}. Realistic imperfect SIC was investigated in \cite{SIC3}, while different power allocation algorithms have been proposed in \cite{wenqiang}. As a future advance, the optimal power allocation maximizing the achievable sum rate was determined in \cite{powerallocation2}, while user scheduling relying on a low-complexity suboptimal approach was conceived in \cite{powerallocation3}. Specifically, the authors of \cite{wenqiang,powerallocation2,powerallocation3} considered NOMA systems using mm-wave carriers which are eminently suitable for ISaC networks. Thus, the investigation of NOMA-based ISaC networks is promising.

\subsubsection{Related literature of ISaC}

The fundamental designs of ISaC networks are investigated in \cite{F_ISAC1,F_ISAC2,F_ISAC3,powerallocation4}, including spectrum sharing methods \cite{F_ISAC1,F_ISAC2}, waveform designs \cite{F_ISAC3}, and resource allocation algorithms \cite{powerallocation4}. The hottest topic in ISaC networks is the investigation of multiple-input-multiple-output (MIMO) ISaC networks, including their MIMO-aided transceiver designs \cite{design,Tcom2,Tcom3,Tcom5,Tcom6}, interference exploitation or interference removal \cite{interference1,interference2,Tcom7}, and the subject of multi-user MIMO ISaC networks \cite{MUMIMO}. But again, the performance analysis of NOMA-based ISaC is still in its infancy. Since the MIMO and NOMA techniques use different domains for multiple access, their comparison, combination, and cooperation under the concept of MIMO ISaC networks is warranted. Additionally, several authors investigated the average performance of ISaC systems relying on the SIC scheme \cite{ref,small}, demonstrating the feasibility of NOMA in Semi-ISaC networks. Hence, there is ample inspiration to pave the way for the evolution of Semi-ISaC networks from OMA to NOMA.

\subsection{Motivation and Contributions}

Again, to consider a practical scenario having high BE, we advocate Semi-ISaC networks, where the bandwidth is split into three portions, namely the communication-only bandwidth, the radar-echo-only bandwidth, and the ISaC bandwidth. Since the NOMA and ISaC concepts match each other harmoniously, we commence by investigating OMA-based Semi-ISaC networks first and then evolve it to a NOMA-based scenario. Our main contributions are summarized as follows:
\begin{itemize}
 \item We propose the novel philosophy of Semi-ISaC networks, where the total bandwidth is split into three portions: the communication-only bandwidth, the radar-echo-only bandwidth, and the ISaC bandwidth. We define both the OMA-based and the NOMA-based Semi-ISaC scenarios. We define three parameters ($\alpha_{semi}$, $\beta_{semi}$, and $ \epsilon_{semi}$) for controlling the bandwidth exploitation of different scenarios.
 \item We evaluate the performance of the OMA-based Semi-ISaC network. As for communication signals,we derive the closed-form expressions of the outage probability (OP) and the ergodic rate. As for radar echoes, we characterize the ergodic Radar Estimation Information Rate (REIR).
  \item We also investigate the performance of the NOMA-based Semi-ISaC networks. To obtain tractable derivations and clear insights, we consider a two-user case, including a communication transmitter and a radar target. We first derive the closed-form expressions of the OP for the communication signals (for both the communication transmitter and the radar target). We then derive the ergodic rate of the communication signals. We also derive the closed-form analytical expressions of the ergodic REIR for the radar echoes.
  \item We evaluate the asymptotic performance of the NOMA-based Semi-ISaC network. Based on the asymptotic expressions, we glean some further insights. We first derive the asymptotic expressions of both the OP and of the ergodic REIR. We then derive both the diversity orders of communication signals and the high signal-to-noise ratio (SNR) slopes for characterizing the radar echoes. As for the diversity orders, analytical results indicate that the near-user's diversity order is $m$, which is the parameter of Nakagami-\emph{m} fading channels, while the far-user's diversity gain is zero. As for the high-SNR slopes of the ergodic REIR, we observe that the high-SNR slopes are related both to the radar's duty cycle and to the pulse duration during its transmission from the base station (BS) to the radar target.
  \item Our numerical results illustrate the following conclusions: 1) For the communication signals, increasing the line-of-sight (LoS) component's transmit power enhances the outage performance; 2) For the radar echoes, dense pulses emerging from the BS enhance the performance of radar detection; and 3) The high-SNR slopes of the radar echoes are also related to the radar's duty cycle and pulse duration, separately.
\end{itemize}

\subsection{Organizations}

The paper is organized as follows. In Section II, we introduce the OMA/NOMA-based Semi-ISaC concepts. In Section III, we investigate the performance of OMA-based Semi-ISaC networks. In Section IV, we evaluate the analytical performance of NOMA-based Semi-ISaC networks, including the analytical outage performance, the ergodic rates of communication signals, as well as the ergodic REIR of the radar echoes. In Section V, we evaluate the system's asymptotic performance and provide insights concerning the NOMA-assisted Semi-ISaC networks, including their asymptotic outage performance for the communication signals having diversity gains and the asymptotic ergodic REIR of the radar echoes by relying on the high-SNR slopes. We then present our numerical results in Section VI, followed by our conclusions in Section VII.

\section{System Model}

\begin{figure*}[t]
\centering
\includegraphics[width= 7in]{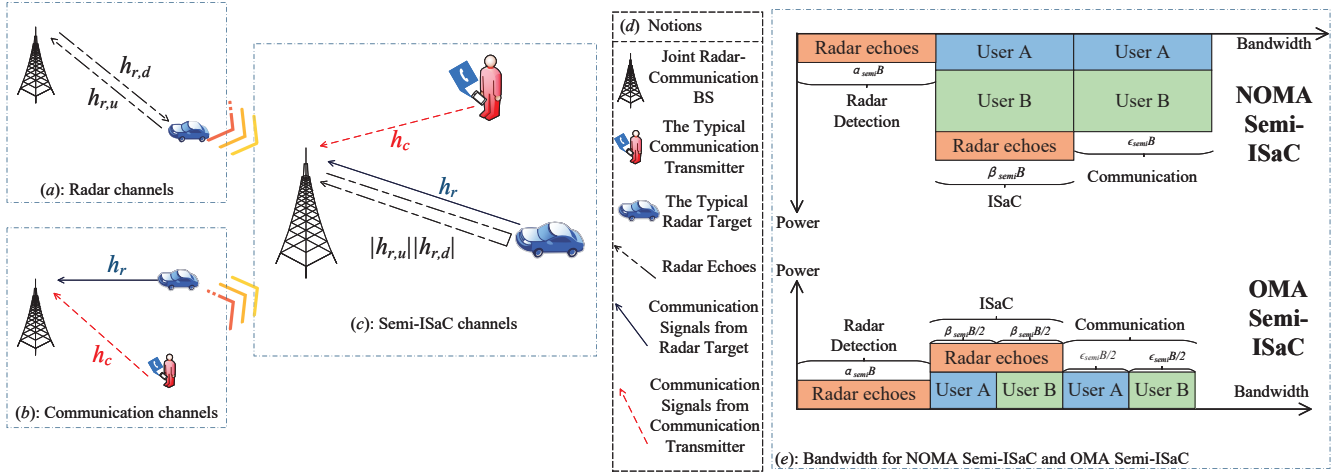}
\caption{Illustration of the NOMA-based Semi-ISaC system: (a) Channels in conventional radar detection systems; (b) Channels in conventional UL NOMA systems; (c) Channels in the NOMA-based Semi-ISaC systems; (d) Notations; and (e) The bandwidth utilization for the NOMA-based Semi-ISaC systems and the OMA-based Semi-ISaC systems;}
\label{system}
\end{figure*}

We focus our attention on an uplink (UL) Semi-ISaC system, which includes a BS, a communication transmitter, and a radar target\footnote{In this paper, we assume that the radar target passively reflects the pulses sent by the BS to indicate the parameters of the radar target, such as range, cross-section, etc. For further information, the radar targets will use the communication function to send signals to the BS in the uplink channels.}. The BS is equipped with an active, mono-static, pulsed radar that exploits the intervals of pulses to detect the radar targets. A single-input-single-output (SISO) model is considered. We assume that the radar target also has communication functions, as exemplified by cars or unmanned aerial vehicles. To design the Semi-ISaC system, the total bandwidth $B$ is split into three portions, including the bandwidth for wireless communication (denoted as $B_W$), the bandwidth for ISaC (denoted as $B_I$), and the bandwidth for radar detection (denoted as $B_R$). We define three coefficients ($\alpha_{semi}$, $\beta_{semi}$, and $ \epsilon_{semi}$) for controlling the bandwidth of different scenarios as:
\begin{align}\label{bandwidth}
B = \underbrace {{\alpha _{semi}}B}_{{B_R}} + \underbrace {{\beta _{semi}}B}_{{B_I}} + \underbrace {{\epsilon_{semi}}B}_{{B_W}}.
\end{align}
where $\alpha_{semi} \in [0,1]$, $\beta_{semi} \in [0,1]$, $\epsilon_{semi} \in [0,1]$, and $\alpha_{semi} + \beta_{semi}+ \epsilon_{semi}=1$.

Before introducing the Semi-ISaC channel model, we highlight our assumptions as follows:
 \begin{itemize}
 \item 1) There is only one radar target located in the serving area of the BS. Other radar targets are served by the BSs of other cells, and hence are ignored in our analysis.
 \item 2) Based on prior observations, the BS is capable of accurately predicting and estimating the time delay of radar echoes to avoid synchronization errors.
 \item 3) The range resolution of the radar system is sufficiently accurate for avoiding the interference between two radar targets\footnote{This assumption is only used for the case when there are more than two radar targets in the serving area of the BS. If there is only a single radar target in the serving area, we do not need this assumption.}.
 \item 4) The range fluctuation is interpreted as a time-delay fluctuation modeled by the Gaussian distribution \cite{ref}.
 \item 5) The cross-section of the radar target is represented by a constant parameter, denoted as $\sigma_{RCS}$. The Doppler shift estimation is perfect for the radar target in order to predict and correct the waveforms.
\end{itemize}

\subsection{Frequency-division (FD) ISaC, OMA-based Semi-ISaC, and NOMA-based Semi-ISaC}

This subsection presents the fundamental concepts and definitions of the conventional (FD) ISaC, OMA-based Semi-ISaC, and NOMA-based Semi-ISaC.
\subsubsection{FD ISaC}
Our benchmark is the FD ISaC having the coefficients of $\alpha_{semi}{=}0$, $\beta_{semi}{=}1$, and $ \epsilon_{semi} {=} 0$. The total bandwidth $B$ is exploited for the ISaC scenario. The users are assigned to orthogonal resource blocks.

\subsubsection{OMA-based Semi-ISaC}
The total bandwidth $B$ is split into three portions with the constraints in Eq. \eqref{bandwidth} as $\alpha_{semi} \in [0,1]$, $\beta_{semi}\in (0,1]$, $ \epsilon_{semi} \in [0,1]$, and $\alpha_{semi}+\beta_{semi}+ \epsilon_{semi}=1$ (which means $\beta_{semi}\ne 0$). In the ISaC bandwidth, SIC is utilized for extracting the radar echo and communication signals from the superimposed signals, while SIC is not utilized to support detecting multiple users with same function. In the following, we consider a two-user case, namely the user A and B. As the radar-echo-only bandwidth ($B_R$) has no communication signal, OMA is used both in the ISaC bandwidth ($B_I$) and in the communication-only bandwidth ($B_W$). Hence, $B_I$ and $B_W$ are further divided into two OMA sub-bands for the two users, respectively. As shown in Fig. \ref{system}. (e), the total bandwidth is finally divided into five parts, including the radar-echo-only bandwidth $B_R$, the ISaC bandwidth for user A $\left(B_I/2\right)$, the ISaC bandwidth for user B $\left(B_I/2\right)$, the communication-only bandwidth for user A $\left(B_W/2\right)$, and the communication-only bandwidth for user B $\left(B_W/2\right)$.

\subsubsection{NOMA-based Semi-ISaC}

As indicated at the top of Fig. \ref{system}. (e), the total bandwidth is split into three portions ($B_W$, $B_I$, and $B_R$) subject to the same coefficient constraints as the OMA-based Semi-ISaC. However, the wireless communication in the bandwidth $B_W$ and $B_I$ relies on NOMA instead of OMA. As exemplified by a two-user case, the communication signals of the two users share the same bandwidth instead of being split into two OMA sub-bands. Additionally, the NOMA-based Semi-ISaC system has to activate SIC in the ISaC bandwidth twice to obtain the radar detection information, while the OMA-based Semi-ISaC scenario only once.

Since the deployment of users directly influence the SIC-based detection orders, we consider two specific deployment scenarios: i) A near communication transmitter is paired with a far radar target, termed as \emph{Scenario-I}; and ii) A far communication transmitter is paired with a near radar target, referred to \emph{Scenario-II}. In the following sections, we evaluate the system performance based on these two scenarios.

\subsection{Channel Model}

\subsubsection{Small-Scale Fading}
The path loss model and small-scale fading model are defined in this subsection for both the radar and communication links. As the ISaC channels are hosted in the mm-wave band, we assume that Nakagami-\emph{m} fading is encountered both by the radar and communication channels \cite{small}. The probability density function (PDF) can be expressed as ${f_{{{\left| {{h_i}} \right|}^2}}}\left( x \right) = \frac{{{m^m}}}{{\Gamma \left( m \right)}}{x^{m - 1}}\exp \left( { - mx} \right)$, with $m$ being the Nakagami-\emph{m} shape parameter and its mean value being one. As seen in Fig. \ref{system}. (a)-(c), the subscript $i = \left\{ {\left( {r,d} \right),\left( {r,u} \right),r,c} \right\}$ represents different small-scale channel gains, where ${\left| {{h_{r,d}}} \right|^2}$ and ${\left| {{h_{r,u}}} \right|^2}$ are those of the downlink (DL) transmission and the UL echo of the radar target, ${{{\left| {{h_c}} \right|}^2}}$ is that of the transmitter's UL communication signal, and ${{{\left| {{h_r}} \right|}^2}}$ is that of the radar target's UL communication signal.

\subsubsection{Large-Scale Fading}
The path loss models of radar echoes are different from that of communication signals. Assume that the distance between the BS and the communication transmitter is $d_c$ and the distance between the BS and the radar target is $d_r$. For the communication channels, the path loss model follows the conventional model of
 \begin{align}\label{pathlosseq}
{\mathcal{P}_c}\left( {{d_c}} \right) = {C_c}{\left( {{d_c}} \right)^{ - {\alpha _c}}},
\end{align}
where $\alpha _c$ is the path loss exponent of communication links, ${C_c}={\left( {\frac{c}{{4\pi {f_c}}}} \right)^2}$ is associated with the reference distance of $d_0 = 1$ m, the speed of light is $c=3\times10^8$ m/s, and the carrier frequency is $f_c$.

We use different coefficients but present the path loss function of the radar echoes in the same form as in Eq. \eqref{pathlosseq}:
\begin{align}
{\mathcal{P}_r}\left( {{d_r}} \right) = {C_r}{\left( {{d_r}} \right)^{ - {\alpha _r}}},
\end{align}
where $\alpha _r$ is the path loss exponent of the radar echoes with $\alpha _r=4$ representing the free-space scenario \cite{small}. The parameter ${C_r} = \frac{{\sigma_{RCS} {\lambda ^2}}}{{{{\left( {4\pi } \right)}^3}}}$ is the reference-distance-based intercept, $\lambda$ is the wavelength of the carrier, and $\sigma_{RCS}= \frac{{4\pi {S_r}}}{{{S_t}}}$ is associated with the target radar cross section, where $S_r$ is the power density that is intercepted by the target and $S_t$ is the scattered power density at the reference distance of $d_0=1$ m \cite{small}.

\subsection{SIC-based Detection Orders for NOMA}

The SIC processes of the conventional NOMA system and the NOMA-based Semi-ISaC system are different. For the conventional NOMA system, the BS only receives signals at two different power levels, when a two-user case is considered, where the near-user receives more power and is detected first compared to the far-user\footnote{As fixed power allocation, the path loss has more dominant effects than the small-scale fading, as we consider the average performance. Thus, we consider the near-user to be the strong user \cite{SIC_large}.}. For the Semi-ISaC NOMA-based system under a two-user case, the BS receives a superposition of various signal components, including the communication signal from the radar target, the communication signal from the communication transmitter, and the radar echo reflected by the radar target. As the BS is capable of estimating the radar echo, we can subtract the estimated radar echo from the superimposed signals to reduce its interference inflicted upon other communication signals. Hence, the communication signals detected from the two users have higher power levels and thus higher priority than the radar echo. As a consequence, it is better to fix the SIC-based detection order of the radar echo to be the last. For the two communication signals, the near-user's signal is detected first and the far-user's signal is detected at the middle stage \cite{ref}.

\subsection{Signal Model}

We aim to support the communication function, but to also add the sensing function into the ISaC system. Since communication signals convey more data than radar pulses over a long period, the communication signals will have high-priority SIC detection orders than the radar echoes. In this case, we fix the radar echo as the last stage of the SIC detection. If the radar echo has a higher power level than those of the communication signals, they will inflict excessive interference. Hence, we subtract a predicted radar echo from the integrated signals to ensure the radar echo's power level is the lowest \cite{ref}. In our model, both OMA and NOMA cases need SIC to split the communication signals and radar pulses. Hence, for both OMA and NOMA cases, subtracting the predicted radar echo may enhance the SIC success rate and then enhance the performance.

We assume that the BS has known the pulse type that was sent to the radar target and acquired prior observations to evaluate the predicted range of the radar target's position. If the radar pulses are regularly sent to the radar target but broadcasting as a fixed frequency, the BS is able to calculate the predicted radar echo based on the prior observations. Naturally, the uncertainty in the positioning directly corresponds to time delay fluctuations in radar systems \cite{ref}. As stated in the assumptions, the time delay fluctuation $\tau$ obeys a Gaussian distribution with the variance of $\sigma_{\tau}^2=\mathbb{E}\left[ {{{\left| {\tau  - {\tau _{pre}}} \right|}^2}} \right]$, where $\mathbb{E}\left[ {\cdot} \right]$ is the expectation. Based on \cite{ref}, we derive the average power level of the radar echo by considering the uncertainty in the positioning decision as:
\begin{align}
\mathbb{E}\left[ {{{\left| {x(t - \tau ) - x(t - {\tau _{pre}})} \right|}^2}} \right] \approx \gamma^2 {\beta_{semi}^2} B^2 \sigma_{\tau}^2,
\end{align}
where we have $\gamma^2 = (2\pi)^2/12$ for a flat spectral shape. The variance $\tau$ is the observation of the time delay for the radar target and $\tau_{pre}$ is the predicted value of $\tau$.

As we fix the SIC-based detection order of the radar echo to be the last, a drawback has to be tolerated, namely that when the radar echo has a high received power level, the ISaC system may face eroded performance, since the radar echo is regarded as interference for the communication signals in the SIC step. To mitigate this, we exploit the predicted target range to generate a predicted radar return and subtract it from the superimposed signals \cite{ref}. We assume that the predicted radar echo is accurate enough for the SIC process. By harnessing this approach, the performance of the communication system is improved. Hence, the received superimposed aggregate signal $v(t)$ is expressed as:
\begin{align}
  &v\left( t \right) = \underbrace {{h_c}\sqrt {{P_c}{\mathcal{P}_c}\left( {{d_c}} \right)} z\left( t \right)}_{{s_c}} + \underbrace {{h_r}\sqrt {{P_r}{\mathcal{P}_c}\left( {{d_r}} \right)} y\left( t \right)}_{{s_r}} \notag\\
  & + \underbrace {{h_{r,d}}{h_{r,u}}\sqrt {{P_{BS}}{\mathcal{P}_r}\left( {{d_r}} \right)} \left[ {x\left( {t - \tau } \right) - x\left( {t - {\tau _{pre}}} \right)} \right]}_{{e_r}} + n\left( t \right),
\end{align}
where $s_c$ is the communication signal received from the UL communication transmitter at the BS, $s_r$ represents the communication signal received from the radar target at the BS, and $e_r$ is the radar echo reflected from the radar target impinging at the BS. Additionally, $P_c$ and $P_r$ are the UL transmit power of the communication transmitter and the radar target, respectively. Moreover, $P_{BS}$ is the UL transmit power of the BS used for radar detection. Finally, $n\left( t \right)$ represents the noise having a variance of $\sigma ^2=k_B T_{temp}\beta_{semi} B$, where $k_B$ is the Boltzmann constant and $T_{temp}$ is the absolute temperature.

Based on the assumptions and derivations above, the signal models of the OMA-based Semi-ISaC and NOMA-based Semi-ISaC are presented in the following part. Additionally, we also summarize the notations of parameters in TABLE I.

\begin{table}
\caption{{Notation of Parameters}}
\renewcommand\arraystretch{1.2}
\label{table_1}
\centering
\newcommand{\tabincell}[2]{\begin{tabular}{@{}#1@{}}#2\end{tabular}}
\begin{tabular}{|p{1.7in}|p{1.4in}|}
\hline
${\mathcal{P}_c}\left( {{d_c}} \right) = {C_c}{\left( {{d_c}} \right)^{ - {\alpha _c}}}$ & ${\mathcal{P}_r}\left( {{d_r}} \right) = {C_r}{\left( {{d_r}} \right)^{ - {\alpha _r}}}$ \\ \hline
${C_c}={\left( {\frac{c}{{4\pi {f_c}}}} \right)^2}$ & ${C_r} = \frac{{\sigma_{RCS} {\lambda ^2}}}{{{{\left( {4\pi } \right)}^3}}}$ \\ \hline
$\Omega  {=} \frac{{m\left( {{P_{BS}}{\mathcal{P}_r}\left( {{d_r}} \right){\gamma ^2}\beta_{semi}^2{B^2}\sigma _\tau ^2 + {\sigma ^2}} \right)}}{{{P_j}{\mathcal{P}_c}\left( {{d_j}} \right)}}$ & ${\Xi _{r,1}} = {2T\beta_{semi}B\gamma _r^{echo}}$ \\ \hline
$a = \frac{2}{{\log _2M}}$ & $b = 2\log _2M\left( {\sin \left( {\frac{\pi }{M}} \right)} \right)$\\ \hline
$Q\left( x \right) = \frac{1}{{\sqrt 2 }}\int_x^\infty  {\exp \left( { - \frac{{{y^2}}}{2}} \right)dy} $ & ${C_n^m}=n!/(m!(n-m)!)$ \\ \hline
${a_1} {=} \frac{{{P_{BS}}{G_r}{C_r}{{\left( {{d_r}} \right)}^{ - {\alpha _r}}}{\gamma ^2}{\beta_{semi}^2}{B^2}\sigma _\tau ^2}}{{{G_c}{C_c}{{\left( {{d_c}} \right)}^{ - {\alpha _c}}}}}$ & ${a_2} {=} \frac{{{\sigma ^2}}}{{{G_c}{C_c}{{\left( {{d_c}} \right)}^{ - {\alpha _c}}}}}$\\ \hline
${a_4} = \frac{{{P_{BS}}{G_r}{C_r}{\gamma ^2}{\beta_{semi}^2}{B^2}\sigma _\tau ^2}}{{{G_c}{C_c}{{\left( {{d_r}} \right)}^{ - {\alpha _c}}}}{{\left( {{d_r}} \right)}^{ {\alpha _r}}}}$ & ${a_3} {=}  \frac{{{{\left( {{d_r}} \right)}^{ - {\alpha _c}}}}}{{{{\left( {{d_c}} \right)}^{ - {\alpha _c}}}}}$\\ \hline
${b_1} = \frac{{{P_{BS}}{G_r}{C_r}{\gamma ^2}{\beta_{semi}^2}{B^2}\sigma _\tau ^2}}{{{G_c}{C_c}{{\left( {{d_r}} \right)}^{{\alpha _r} - {\alpha _c}}}}}$ & ${a_5} = \frac{{{\sigma ^2}}}{{{G_c}{C_c}{{\left( {{d_r}} \right)}^{ - {\alpha _c}}}}}$ \\ \hline
${b_{\text{2}}}{\text{ = }}\frac{{{\sigma ^2}}}{{{G_c}{C_c}{{\left( {{d_r}} \right)}^{ - {\alpha _c}}}}}$ & ${b_{\text{3}}}=\frac{{{{\left( {{d_c}} \right)}^{ - {\alpha _c}}}}}{{{{\left( {{d_r}} \right)}^{ - {\alpha _c}}}}}$ \\ \hline
${\Lambda _1} = \frac{{m\left( {{a_1} + {a_2}} \right)}}{{{P_c}}}$ & ${\Lambda _3} = \frac{{m\left( {{b_{\text{1}}}{\text{  +  }}{b_{\text{2}}}} \right)}}{{{P_r}}}$ \\ \hline
${\Lambda _2} = \frac{{m\left( {{a_4} + {a_5}} \right)}}{{{P_r}}}\left( {\frac{{{\gamma _{SIC}}{a_3}{P_r}}}{{{P_c}}} + 1} \right)$  &  ${\Lambda _5} = \frac{{m{\gamma _{SIC}}\left( {{b_{\text{1}}}{\text{  +  }}{b_{\text{2}}}} \right)}}{{{P_r}}}$  \\ \hline
\end{tabular}
\begin{tabular}{|p{3.272in}|}
${\Lambda _4} = \frac{m}{{{P_c}}}\left( {{a_1} + {a_2}} \right)\left( {\frac{{{\gamma _{SIC}}}}{{{P_r}}}{P_c}{b_{\text{3}}} + 1} \right)$
\\ \hline
\end{tabular}
\end{table}

\subsubsection{Communication Signals for OMA-based Semi-ISaC}
The signal-to-interference-and-noise ratio (SINR) expression of the communication transmitter and the radar target is expressed as:
\begin{align}\label{S-OMA}
\gamma _j^{OMA} = \frac{{{P_j}{{\cal P}_c}\left( {{d_j}} \right){{\left| {{h_j}} \right|}^2}}}{{{P_{BS}}{{\cal P}_r}\left( {{d_r}} \right){{\left| {{g_r}} \right|}^2} + {\sigma ^2}}},
 \end{align}
where $j \in \{c,r\}$ represents for the communication transmitter and the radar target, respectively. The channel fading parameter is denoted as \begin{align}
{\left| {{g_r}} \right|^2} = {\left| {{h_{r,d}}} \right|^2}{\left| {{h_{r,u}}} \right|^2}{\gamma ^2}\beta _{semi}^2{B^2}\sigma _\tau ^2.
 \end{align}
\subsubsection{Communication Signals for NOMA-based Semi-ISaC in Scenario-I}

For Scenario-I, the communication transmitter is the near-user, whose signals is detected first. Given the different power levels, the BS directly detects the UL signal received from the communication transmitter by considering both the communication signals and the radar echo of the radar target as interference. Hence, the SINR of the communication transmitter is formulated as:
\begin{align}\label{S-I-c}
\gamma _c^I = \frac{{\overbrace {{P_c}{{\cal P}_c}\left( {{d_c}} \right){{\left| {{h_c}} \right|}^2}}^{{\text{Transmitter's Communication Signals}}}}}{{\underbrace {{P_r}{{\cal P}_c}\left( {{d_r}} \right){{\left| {{h_r}} \right|}^2}}_{\scriptstyle{\text{Radar Target's }}\hfill\atop
\scriptstyle{\text{Communication Signals}}\hfill} + \underbrace {{P_{BS}}{{\cal P}_r}\left( {{d_r}} \right){{\left| {{g_r}} \right|}^2}}_{{\text{Radar Echoes}}} + \underbrace {{\sigma ^2}}_{{\text{Noise}}}}}.
 \end{align}

By subtracting the signal of the communication transmitter from the composite signal by the SIC remodulated process, the SINR of the communication signals for the radar target becomes
\begin{align}\label{S-I-r}
\gamma _r^I = \frac{{\overbrace {{P_r}{{\cal P}_c}\left( {{d_r}} \right){{\left| {{h_r}} \right|}^2}}^{{\text{Radar Target's Communication Signals}}}}}{{\underbrace {{\varsigma _c}{P_c}{{\cal P}_c}\left( {{d_c}} \right){{\left| {{h_c}} \right|}^2}}_{\scriptstyle{\text{SIC of Transmitter's}}\hfill\atop
\scriptstyle{\text{Communication Signals}}\hfill} + \underbrace {{P_{BS}}{{\cal P}_r}\left( {{d_r}} \right){{\left| {{g_r}} \right|}^2}}_{{\text{Radar Echoes}}} + \underbrace {{\sigma ^2}}_{{\text{Noise}}}}},
 \end{align}
where $0< {{\varsigma _c}}<1 $ represents the imperfect SIC while ${{\varsigma _c}}=0 $ corresponds to the perfect SIC.

\subsubsection{Communication Signals for NOMA-based Semi-ISaC in Scenario-II}

For Scenario-II, the near-user is the radar target. Thus, the BS firstly detects the communication signals of the radar target, yielding an SINR of
\begin{align}\label{S-II-r}
\gamma _r^{II} = \frac{{\overbrace {{P_r}{{\cal P}_c}\left( {{d_r}} \right){{\left| {{h_r}} \right|}^2}}^{{\text{Radar Targets's Communication Signals}}}}}{{\underbrace {{P_c}{{\cal P}_c}\left( {{d_c}} \right){{\left| {{h_c}} \right|}^2}}_{\scriptstyle{\text{Transmitter's }}\hfill\atop
\scriptstyle{\text{Communication Signals}}\hfill} + \underbrace {{P_{BS}}{{\cal P}_r}\left( {{d_r}} \right){{\left| {{g_r}} \right|}^2}}_{{\text{Radar Echoes}}} + \underbrace {{\sigma ^2}}_{{\text{Noise}}}}}.
\end{align}

Following the (perfect/imperfect) SIC process, the SINR of communication signals for the communication transmitter becomes
\begin{align}\label{S-II-c}
\gamma _c^{II} = \frac{{\overbrace {{P_c}{{\cal P}_c}\left( {{d_c}} \right){{\left| {{h_c}} \right|}^2}}^{{\text{Transmitter's Communication Signals}}}}}{{\underbrace {{\varsigma _r}{P_r}{{\cal P}_c}\left( {{d_r}} \right){{\left| {{h_r}} \right|}^2}}_{\scriptstyle{\text{SIC of Radar Targets's }}\hfill\atop
\scriptstyle{\text{Communication Signals}}\hfill} + \underbrace {{P_{BS}}{{\cal P}_r}\left( {{d_r}} \right){{\left| {{g_r}} \right|}^2}}_{{\text{Radar Echoes}}} + \underbrace {{\sigma ^2}}_{{\text{Noise}}}}},
\end{align}
where $0< {{\varsigma _r}}<1 $ represents the imperfect SIC while ${{\varsigma _r}}=0 $ corresponds to the perfect SIC.

\subsubsection{Radar Echoes for OMA and NOMA}

Since we aim to ensure the priority of communication signals, the radar echo is simply left behind after the last SIC stage. With the aid of SIC, the SNR is expressed as:
\begin{align}
\gamma _r^{echo} = \frac{{\overbrace {{P_{BS}}{{\cal P}_r}\left( {{d_r}} \right){{\left| {{g_r}} \right|}^2}}^{{\text{Radar Echoes}}}}}{{\underbrace {{\varsigma _c}{P_c}{{\cal P}_c}\left( {{d_c}} \right){{\left| {{h_c}} \right|}^2} + {\varsigma _r}{P_r}{{\cal P}_c}\left( {{d_r}} \right){{\left| {{h_r}} \right|}^2}}_{{\text{SIC of Communication Signals}}} + {\sigma ^2}}}.
\end{align}
For the equation above, both the NOMA and OMA cases associated with perfect SIC have ${{\varsigma _c}}=0$ and ${{\varsigma _r}}=0$. The NOMA case with imperfect SIC has $0<{{\varsigma _c}}<1$ and $0<{{\varsigma _r}}<1$. The OMA case with imperfect SIC has two combinations: 1) $0<{{\varsigma _c}}<1$ and ${{\varsigma _r}}=0$ for the communication transmitter's subchannel and 2) ${{\varsigma _c}}=0$ and $0<{{\varsigma _r}}<1$ for the radar target's subchannel.

In Sections III to V, we will consider the ergodic REIR as the metric for characterizing the performance of the radar detection system. This metric is directly related to $\gamma_r^{echo}$ derived above.

\subsubsection{Perfect or Imperfect SIC}

This paper aims to first propose the Semi-ISaC network, hence we exploit perfect SIC schemes to investigate the performance of upper bounds (${{\varsigma _c}}=0$ and ${{\varsigma _r}}=0$). Based on the derivations in Sections III to V, we could have some insights to indicate the properties of the Semi-ISaC network. As for the imperfect SIC scenarios, we will draw a figure in Section VI to compare the performance between the upper bounds and practical scenarios. The analytical derivation and investigation of imperfect SIC cases can be extended by our model and will be left for our future research due to the strict limitation of space.

\section{Performance Evaluation for OMA-based Semi-ISaC}
In this section, we evaluate the OMA-based Semi-ISaC networks. Again, we adopt the OP and the ergodic rate as the performance metrics for communication signals. Likewise, the ergodic REIR is adopted as the performance metric for the radar echoes.

\subsection{Performance Evaluation for Communication Signals}

In this subsection, we aim to investigate the performance of communication signals. Before that, we first evaluate the average interference strength.

\begin{lemma}\label{AveragedInterference}
\emph{To simplify the expression of interference (radar echoes), we introduce the shorthand of ${I_R} = {P_{BS}}{\mathcal{P}_r}\left( {{d_r}} \right){\left| {{h_{r,d}}} \right|^2}{\left| {{h_{r,u}}} \right|^2}{\gamma ^2}{B^2}\sigma _\tau ^2$. The expectation of interference is expressed as: }
\begin{align}
\mathbb{E}\left[ {{I_R}} \right]\left( {{d_r}} \right) = {P_{BS}}{\mathcal{P}_r}\left( {{d_r}} \right){\gamma ^2}{\beta_{semi}^2}{B^2}\sigma _\tau ^2.
\end{align}

Sketch of Proof:
Given the definition of expectation and the distribution of Nakagami-m fading channels, the expression of interference is presented as:
\begin{align}
&\mathbb{E}\left[ {{I_R}} \right]\left( {{d_r}} \right) = {P_{BS}}{\mathcal{P}_r}\left( {{d_r}} \right){\gamma ^2}{\beta_{semi}^2}{B^2}\sigma _\tau ^2{\left( {\frac{{{m^m}}}{{\Gamma \left( m \right)}}} \right)^2}\notag\\
&\hspace*{0.2cm}\times \int_0^\infty  {{x^m}\exp \left( { - mx} \right)dx} \int_0^\infty  {{y^m}\exp \left( { - my} \right)} dy,
\end{align}
and with the aid of Eq. [2.3.3.1] in \cite{table}, this lemma is proved. We have the detailed proof in Section I of \cite{proof}.
\end{lemma}

In the OMA-based Semi-ISaC network, the OP of the communication signals is defined as $\mathbb{P}_j^{OMA} = \Pr \left\{ {\gamma _j^{OMA} < \gamma _{th}^{OMA}} \right\}$, given the threshold $\gamma _{th}^{OMA}$. The achieved rate is defined as $R_j^{OMA} = \frac{1}{2}{\log _2}$ $\left( {1 + \gamma _j^{OMA}} \right)$. \textbf{Theorem \ref{OPEROMA}} provides the closed-form expressions of both the OP and the ergodic rate for communication signals in the OMA-based Semi-ISaC network.

\begin{theorem}\label{OPEROMA}
\emph{Upon introducing the subscript of $j \in \{c,r\}$ for representing the communication transmitter and the radar target, the expression of the OP and that of the ergodic rate are derived respectively as:}
\begin{align}
\mathbb{P}_j^{OMA}  &= \frac{{\gamma \left( {m,\Omega \gamma _{th}^{OMA}} \right)}}{{\Gamma \left( m \right)}}, \\
R_j^{OMA}  &= \frac{1}{{2\ln 2}}\sum\limits_{k = 0}^{m - 1} {\exp \left( \Omega  \right){E_{1 + k}}\left( \Omega  \right)} ,
\end{align}
\emph{where we have $\Omega  {=} \frac{{m\left( {{P_{BS}}{\mathcal{P}_r}\left( {{d_r}} \right){\gamma ^2}\beta_{semi}^2{B^2}\sigma _\tau ^2 + {\sigma ^2}} \right)}}{{{P_j}{\mathcal{P}_c}\left( {{d_j}} \right)}}$, $\Gamma(x)$ is the Gamma function, $\gamma(a,b)$ is the incomplete Gamma function, and ${E_n}\left( \cdot \right)$ is the generalized exponential integral.}

Sketch of Proof: We derive the OP by exploiting the cumulative distribution function (CDF) of the Gamma distribution, denoted as ${F_{{{\left| {{h_j}} \right|}^2}}}\left( x \right) = \frac{{\gamma \left( {m,mx} \right)}}{{\Gamma \left( m \right)}}$. We additionally derive the ergodic rate by exploiting $\gamma \left( {m,t} \right) = \left( {m - 1} \right)! - \exp \left( { - t} \right)\sum\limits_{k = 0}^{m - 1} {\frac{{\left( {m - 1} \right)!}}{{k!}}{t^k}} $, $\Gamma \left( { - k,\Omega } \right) = \frac{{{E_{1 + k}}\left( \Omega  \right)}}{{{\Omega ^k}}}$, and $\int_0^\infty  {\frac{{{x^a}}}{{1 + x}}} \exp \left( { - bx} \right) = \exp \left( b \right)\Gamma \left( {a + 1} \right)\Gamma \left( { - a,b} \right)$, where $\Gamma(a,b)$ is the incomplete Gamma function. The detailed proof is presented in Section II of \cite{proof}.

\end{theorem}

\subsection{Performance Evaluation for Radar Echoes}

Again for radar echoes, the authors of \cite{ref} have proposed the REIR metric to evaluate the performance of radar targets. The REIR is analogous to the data information rate of the communications system. This is the calculated estimation rate of the parameters (range, cross-section, etc.). A higher REIR means better performance for radar detection. We represents a clear relationship between the REIR and the SNR $\gamma _r^{echo}$, presented as:
\begin{align}\label{R_est}
{R_{est}} \leqslant \frac{\delta }{{2T}}{\log _2}\left( {1 + 2T{\beta_{semi}}B\gamma _r^{echo}} \right),
\end{align}
where $\gamma _r^{echo} = \frac{{{P_{BS}}{\mathcal{P}_r}\left( {{d_r}} \right){{\left| {{h_{r,d}}} \right|}^2}{{\left| {{h_{r,u}}} \right|}^2}{\gamma ^2}{\beta_{semi}^2}{B^2}\sigma _\tau ^2}}{{{\sigma ^2}}}$ is the SNR for the radar echoes of the radar target, $T$ is the radar pulse duration, and $\delta$ is the radar's duty cycle. We then use the ergodic REIR for quantifying the average radar estimation rate, which may be viewed as the dual counterpart of the data information rate, presented as:
\begin{align}\label{ER_est}
{R_{est}} \leqslant \mathbb{E}\left[\frac{\delta }{{2T}}{\log _2}\left( {1 + 2T{\beta_{semi}}B\gamma _r^{echo}} \right) \right].
\end{align}

The following expression shows the relationship between the REIR and radar estimation. We note that the time-delay estimation is a basic range measurement, denoted as $\sigma_{\tau, est}^2$. Our REIR metric is characterized by the Cram\'{e}r-Rao lower bound (CRLB) of the radar estimation (range measurement)\cite{Tcom1,Tcom4}, denoted as $\sigma_{\tau, est}^2=\frac{\sigma_{\tau }^2}{2 T \beta_{semi} B \gamma _r^{echo}}$. First, we have the definition of the REIR is $R_{est} \leq \frac{H_{\tau_{\mathrm{rr}}} - H_{\tau_{\mathrm{est}}}} {T_{\mathrm{bit}}}$, where $H_{\tau_{\mathrm{rr}}}$ is the entropy of received signal, denoted as $H_{\tau_{\mathrm{rr}}} = \frac{1}{2} \log _2\left[2 \pi e\left(\sigma_{\tau }^2+\sigma_{\tau, est}^2\right)\right]$, $H_{\tau_{\mathrm{est}}}$ is the entropy of errors, denoted as $H_{\tau_{\mathrm{est}}} = \frac{1}{2} \log_2 \left[2 \pi e\sigma_{\tau, est}^2\right]$, and ${T_{\mathrm{bit}}} = T/\delta$ represents the bits per pulse repetition interval. Hence, we could derive the REIR by substituting the CRLB into the definition equation. We also conclude that the REIR is strongly influenced by the radar's time-delay estimation.

\subsubsection{Equivalent Radar Channels}
The radar channel may be considered as a pair of independent serially concatenated links, constituted by the DL channel spanning from the BS to the radar target and the UL channel reflected from the radar target back to the BS. Thus, the equivalent small-scale channel gain may be expressed by ${\left| {{h_{r,eq}}} \right|^2} = {\left| {{h_{r,d}}} \right|^2}{\left| {{h_{r,u}}} \right|^2}$. We first derive the distribution of ${\left| {{h_{r,eq}}} \right|^2}$ in \textbf{Lemma \ref{channel_lemma}} and the ergodic REIR is then given in \textbf{Theorem \ref{EREIR1}}.

\begin{lemma}\label{channel_lemma}
\emph{If the UL and DL channels are Nakagami-\emph{m} fading channels, the PDF and CDF of the equivalent channel gain is expressed as:}
\begin{align}
  {f_{{{\left| {{h_{r,eq}}} \right|}^2}}}\left( z \right) &= \frac{{2{m^{2m}}}}{{{{\left( {\Gamma \left( m \right)} \right)}^2}}}{z^{m - 1}}{K_0}\left( {2m\sqrt z } \right) ,\\
  {F_{{{\left| {{h_{r,eq}}} \right|}^2}}}\left( z \right) &= \frac{{G{_1^2}{_3^1}\left( {{m^2}x\left| {_{m,m,0}^1} \right.} \right)}}{{{{\left( {\Gamma \left( m \right)} \right)}^2}}} ,
\end{align}
\emph{where ${K_0}\left(  \cdot  \right)$ is the modified Bessel function of the third kind and ${G{_p^m}{_q^n}\left( {\cdot \left| {_{\left( {{b_q}} \right)}^{\left( {{a_p}} \right)}} \right.} \right)}$ is the Meijer G function.}

Sketch of Proof: We derive the above PDF and CDF by noticing ${K_v}\left( x \right) = \frac{1}{2}G{_0^2}{_2^0}\left( {\frac{{{x^2}}}{4}\left| {{_{\frac{v}{2}}^ \cdot}{ {_{\frac{ - v}{2}}^ \cdot}} } \right.} \right)$, $ {z^p}G{_p^m}{_q^n}\left( {z\left| {_{\left( {{b_q}} \right)}^{\left( {{a_p}} \right)}} \right.} \right) = G{_p^m}{_q^n}\left( {z\left| {_{\left( {{b_q}} \right) + p}^{\left( {{a_p}} \right) + p}} \right.} \right)$, $\int_0^x {{z^{m - 1}}G{_0^2}{_2^0}\left( {{m^2}z\left| {{_0^ \cdot}{ _0^ \cdot} } \right.} \right)} dz = {x^m}G{_1^2}{_3^1}\left( {{m^2}y\left| {{_{0,0}^{1 - m}}{_{ - m}^ \cdot} } \right.} \right)$, and Eq.[2.3.6.7] in \cite{table}. We present the comprehensive proof in Section III of \cite{proof}.
\end{lemma}

\subsubsection{Ergodic REIR}

Based on the equivalent channel distribution, we will derive the ergodic REIR of the radar echoes in \textbf{Theorem \ref{EREIR1}}. We will also exploit \textbf{Corollary \ref{EREIR3}} to evaluate the performance under the Rayleigh fading channels.
\begin{theorem}\label{EREIR1}
\emph{For the analytical results of the radar echoes, the expressions of the ergodic REIR are formulated as:}
\begin{align}\label{R_r_low}
{R_{est}^{low}} {=} \frac{\delta }{{2T\ln \left( 2 \right)}}\int_0^\infty  {\frac{1}{{z {+} 1}}\left( {1 {-} \frac{{G{_1^2}{_3^1}\left( {\frac{{{m^2}d_t^{{\alpha _r}}}}{{{\Xi _{r,1}}}}z\left| {_{m,m,0}^1} \right.} \right)}}{{{{\left( {\Gamma \left( m \right)} \right)}^2}}}} \right)} dz,
\end{align}
\emph{where we have ${\Xi _{r,1}} = {2T\beta_{semi}B\gamma _r^{echo}}$.}

Sketch of Proof:
With the aid of \textbf{Lemma \ref{channel_lemma}}, this theorem is proved. The comprehensive proof is presented in Section IV of \cite{proof}.
\end{theorem}

\begin{corollary}\label{EREIR3}
\emph{Assuming that the radar channel experiences Raleigh fading, the ergodic rate in Eq. \eqref{R_r_low} is simplified as: }
\begin{align}
{R_{est}^{low}} = \frac{\delta }{{2T\ln \left( 2 \right)}}G{_1^3}{_3^1}\left( {d_t^{{\alpha _r}}\Xi _{r,1}^{ - 1}\left| {_{0,0,1}^0} \right.} \right).
\end{align}

Sketch of Proof: This corollary is proved by exploiting Eq.[2.3.4.4] in \cite{table} and the Meijer G function. Detailed derivations are similar to those of \textbf{Lemma \ref{channel_lemma}}. The comprehensive proof is jointly presented in Section IV of \cite{proof} with \textbf{Theorem \ref{EREIR1}}.
\end{corollary}

\begin{remark}
The ergodic REIR insightfully characterizes the performance of the radar detection. We still leave more open space for other metrics to represent the performance of the radar detection, such as bit error rate (BER). For example, under M-PSK, we could express the BER expression as
\begin{align}
{\varepsilon _{{\text{BER}}}} = a_{BER}Q\left( {\sqrt {b_{BER}\gamma _r^{echo}} } \right),
\end{align}
where $a_{BER} = \frac{2}{{\log _2M}}$, $b_{BER} = 2\log _2M\left( {\sin \left( {\frac{\pi }{M}} \right)} \right)$, and $Q\left( x \right) = \frac{1}{{\sqrt 2 }}\int_x^\infty  {\exp \left( { - \frac{{{y^2}}}{2}} \right)dy} $ \cite{BER}.
\end{remark}

\section{Analytical Performance Evaluation for NOMA-based Semi-ISaC}

In this section, we analyze the performance metrics for NOMA-based Semi-ISaC networks. The analytical results in this section will be useful in Section V to obtain deep insights.

\subsection{Performance Analysis for Communication Signals in Scenario-I}

Recall that the communication transmitter is the near-user and the radar target is the far-user in Scenario-I. The OP expressions for the NOMA users in Scenario-I are given by
\begin{align}
\label{OP-I-c}
\mathbb{P}_c^I &= \Pr \left\{ {\gamma _c^I < {\gamma _{th}}} \right\},\\
\label{OP-I-r}
\mathbb{P}_r^I &= 1 - \Pr \left\{ {\gamma _c^I > {\gamma _{SIC}},\gamma _r^I > {\gamma _{th}}} \right\},
\end{align}
where $\Pr \left\{ \mathcal{A},\mathcal{B}\right\}$ is the probability that both $\mathcal{A}$ and $\mathcal{B}$ are true, $\gamma _{SIC}$ is the threshold of the SIC process, and $\gamma_{th}$ is the threshold of communication signal transmission in the NOMA-based Semi-ISaC scenario. If the OP is lower than the threshold, the communication fails and vise versa.

In the following, the closed-form expressions of the OP and the ergodic rate for a pair of NOMA users are given in \textbf{Theorem \ref{OP_I_c_f}-\ref{OP_I_r_f}} and \textbf{Corollary \ref{OP_I_c_r}-\ref{OP_I_r_r}}.

\begin{theorem}\label{OP_I_c_f}
\emph{In Scenario-I of the NOMA-based Semi-ISaC scenario, the OP expression of the communication transmitter is }
\begin{align} \label{Afinal}
  &\mathbb{P}_c^I = 1 - \exp \left( { - \frac{{m{\gamma _{th}}}}{{{P_c}}}\left( {{a_1} + {a_2}} \right)} \right)\sum\limits_{p = 0}^{m - 1} {\frac{{{m^r}\gamma _{_{th}}^p}}{{\left( {m - 1} \right)!p!}}}  \notag\\
  & \hspace*{0.1cm} \times \sum\limits_{r = 0}^p {C_p^r \Gamma \left( {m + p - r} \right) \frac{{{{\left( {{a_1} + {a_2}} \right)}^r}{{\left( {{P_r}{a_3}} \right)}^{p - r}}}}{{P_c^p}{\left( {\frac{{{\gamma _{th}}{a_3}{P_r}}}{{{P_c}}} + 1} \right)^{m + p - r}}}}  ,
\end{align}
\emph{where we have ${a_1} = \frac{{{P_{BS}}{G_r}{C_r}{{\left( {{d_r}} \right)}^{ - {\alpha _r}}}{\gamma ^2}{\beta_{semi}^2}{B^2}\sigma _\tau ^2}}{{{G_c}{C_c}{{\left( {{d_c}} \right)}^{ - {\alpha _c}}}}}$, ${a_2} = \frac{{{\sigma ^2}}}{{{G_c}{C_c}{{\left( {{d_c}} \right)}^{ - {\alpha _c}}}}}$, ${a_3} =  \frac{{{{\left( {{d_r}} \right)}^{ - {\alpha _c}}}}}{{{{\left( {{d_c}} \right)}^{ - {\alpha _c}}}}}$, and ${C_n^m}=n!/(m!(n-m)!)$. }

Sketch of Proof:
See Appendix~A.
\end{theorem}

\begin{corollary}\label{OP_I_c_r}
\emph{In Scenario-I, the ergodic rate of the communication transmitter in the NOMA-based Semi-ISAC scenario is formulated as:}
\begin{align}
  & R_c^{er,I} = \frac{1}{{\ln 2}}\sum\limits_{p = 0}^{m - 1} {\sum\limits_{r = 0}^p {C_p^r\frac{{\Lambda _1^{r - \left( {1 + p + k} \right)}{{\left( {{P_r}{a_3}} \right)}^{p - r}}}}{{\left( {m - 1} \right)!p!P_c^{p - r}}}} } \notag \\
  &\hspace*{0.1cm} \times \Gamma \left( {m + p - r} \right)\sum\limits_{k = 0}^\infty  \binom{m + p - r + k - 1 }{k}  \notag \\
  &\hspace*{0.1cm} \times {\left( { - \frac{{{a_3}{P_r}}}{{{P_c}}}} \right)^k}\exp \left( {{\Lambda _1}} \right)\Gamma \left( {p + k + 1} \right){E_{1 + p + k}}\left( {{\Lambda _1}} \right),
\end{align}
\emph{where we have ${\Lambda _1} = \frac{{m\left( {{a_1} + {a_2}} \right)}}{{{P_c}}}$.}

Sketch of Proof:
By substituting the equation in \textbf{Theorem \ref{OP_I_c_f}} into the definition of the ergodic rate, which is expressed as $R_c^{er,I} = \frac{1}{{\ln 2}}\int_0^\infty  {\frac{{1 - \mathbb{P}_c^I\left( {{\gamma _{th}}} \right)}}{{1 + {\gamma _{th}}}}} d{\gamma _{th}}$, the ergodic rate expression is given by
\begin{align}
  &R_c^{er,I} = \frac{1}{{\ln 2}}\sum\limits_{p = 0}^{m - 1} {\frac{{{m^r}}}{{\left( {m - 1} \right)!p!}}} \sum\limits_{r = 0}^p {C_p^r\frac{{{{\left( {{a_1} + {a_2}} \right)}^r}{{\left( {{P_r}{a_3}} \right)}^{p - r}}}}{{P_c^p}}}  \notag \\
 &\hspace*{0.1cm}\times  \Gamma \left( {m + p - r} \right)\int_0^\infty  {\frac{{{x^p}}}{{1 + x}}} {\left( {\frac{{x{a_3}{P_r}}}{{{P_c}}} + 1} \right)^{ - \left( {m + p - r} \right)}}\notag\\
 &\hspace*{0.1cm} \times \exp \left( { - \frac{{m{x}}}{{{P_c}}}\left( {{a_1} + {a_2}} \right)} \right)dx.
\end{align}
The corollary can be proved by noting ${\left( {1 + x} \right)^{ - n}} = \sum\limits_{k = 0}^\infty  {\binom{  n + k - 1 }{  k }} {\left( { - x} \right)^k}$, $\Gamma \left( {a,b} \right) = \sum\limits_{p = 0}^{a - 1} \frac{{\left( {a - 1} \right)!}}{{p!}}{b^p}$ $\times\exp \left( { - b} \right) $, ${E_n}\left( x \right) = {x^n}\Gamma \left( {1 - n,x} \right)$, and $\int_0^\infty  {\frac{{{x^a}}}{{1 + x}}} \exp \left( { - bx} \right) = \exp \left( b \right)\Gamma \left( {a + 1} \right)\Gamma \left( { - a,b} \right)$. We have the detailed proof in Section V of \cite{proof}.
\end{corollary}

\begin{theorem}\label{OP_I_r_f}
\emph{In Scenario-I of NOMA-based Semi-ISaC, the OP of the radar target is given by }
\begin{align} \label{Bfinal}
  &\mathbb{P}_r^I =1 - \sum\limits_{p = 0}^{m - 1} {\sum\limits_{r = 0}^p {C_p^r} \frac{{{{\left( {{a_1} + {a_2}} \right)}^{p - r}}{{\left( {{a_3}{P_r}} \right)}^r}}}{{\Gamma \left( m \right){m^r}p!}}} {\left( {\frac{{m{\gamma _{SIC}}}}{{{P_c}}}} \right)^p}\notag\\
&\hspace*{0.1cm} \times \exp \left( { - \frac{{m{\gamma _{SIC}}\left( {{a_1} + {a_2}} \right)}}{{{P_c}}}} \right){\left( {\frac{{{\gamma _{SIC}}{a_3}{P_r}}}{{{P_c}}} + 1} \right)^{ - (r + m)}}\notag\\
& \hspace*{0.1cm}\times \Gamma \left( {r + m,\frac{{{\gamma _{th}}m\left( {{a_4} + {a_5}} \right)}}{{{P_r}}}\left( {\frac{{{\gamma _{SIC}}{a_3}{P_r}}}{{{P_c}}} + 1} \right)} \right),
\end{align}
\emph{where we have ${a_4} = \frac{{{P_{BS}}{G_r}{C_r}{{\left( {{d_r}} \right)}^{ - {\alpha _r}}}{\gamma ^2}{\beta_{semi}^2}{B^2}\sigma _\tau ^2}}{{{G_c}{C_c}{{\left( {{d_r}} \right)}^{ - {\alpha _c}}}}}$ and ${a_5} = \frac{{{\sigma ^2}}}{{{G_c}{C_c}{{\left( {{d_r}} \right)}^{ - {\alpha _c}}}}}$.}

Sketch of Proof:
See Appendix~B.
\end{theorem}

\begin{corollary}\label{OP_I_r_r}
\emph{In Scenario-I, the ergodic rate expression for the communication signal of the radar target is derived as:  }
\begin{align}
  &R_r^{er,I} = \frac{1}{{\ln 2}}\sum\limits_{p = 0}^{m - 1} {\sum\limits_{r = 0}^p {C_p^r} \frac{{{{\left( {{a_1} + {a_2}} \right)}^{p - r}}{{\left( {{a_3}{P_r}} \right)}^r}}}{{\Gamma \left( m \right){m^r}p!}}{{\left( {\frac{{m{\gamma _{SIC}}}}{{{P_c}}}} \right)}^p}} \notag\\
&\hspace*{0.1cm} \times \exp \left( { - {\Lambda _1}{\gamma _{SIC}}} \right){\left( {\frac{{{\gamma _{SIC}}{a_3}{P_r}}}{{{P_c}}} + 1} \right)^{ - (r + m)}}\sum\limits_{k = 0}^{r + m - 1} {\frac{{{\Lambda _2}}}{{k!}}} \notag\\
& \hspace*{0.1cm}\times \left( {r + m - 1} \right)!\exp \left( {{\Lambda _2}} \right)\Gamma \left( {k + 1} \right){E_{1 + k}}\left( {{\Lambda _2}} \right),
\end{align}
\emph{where we have ${\Lambda _2} = \frac{{m\left( {{a_4} + {a_5}} \right)}}{{{P_r}}}\left( {\frac{{{\gamma _{SIC}}{a_3}{P_r}}}{{{P_c}}} + 1} \right)$.}

Sketch of Proof:
The proof is similar to that of \textbf{Theorem \ref{OPEROMA}}.
\end{corollary}

\subsection{Performance Analysis for Communication Signals in Scenario-II}

This subsection evaluates both the OP and the ergodic rate of NOMA-based Semi-ISaC in Scenario-II. Compared to Scenario-I, the SIC detection orders are the opposite way round. Thus, the OP expressions become
\begin{align}
\label{OP-II-r}
\mathbb{P}_r^{II}& = \Pr \left\{ {\gamma _r^{II} < {\gamma _{th}}} \right\},\\
\label{OP-II-c}
\mathbb{P}_c^{II} & = 1 - \Pr \left\{ {\gamma _r^{II} > {\gamma _{SIC}},\gamma _c^{II} > {\gamma _{th}}} \right\},
\end{align}
and the expressions of the OP and those of the ergodic rate are presented by \textbf{Theorem \ref{OP_II_r_f}-\ref{OP_II_c_f}} and \textbf{Corollary \ref{OP_II_r_r}-\ref{OP_II_c_r}}.

\begin{theorem}\label{OP_II_r_f}
\emph{For NOMA-based Semi-ISaC in Scenario-II, the OP for the communication signal of the radar target is formulated as:  }
\begin{align}
 & \mathbb{P}_r^{II} = 1 - \sum\limits_{p = 0}^{m - 1} {\sum\limits_{r = 0}^p {\frac{{C_p^r\Gamma \left( {m + r} \right)}}{{\Gamma \left( m \right){m^r}p!}}} } {\left( {\frac{{m{\gamma _{th}}}}{{{P_r}}}} \right)^p}\notag\\
 &\hspace*{0.1cm} \times \exp\left( { - \frac{{m{\gamma _{th}}\left( {{b_{\text{1}}}{\text{ + }}{b_{\text{2}}}} \right)}}{{{P_r}}}} \right) \notag\\
 &\hspace*{0.1cm}\times  {\left( {{b_{\text{1}}}{\text{ + }}{b_{\text{2}}}} \right)^{p - r}}{\left( {{P_c}{b_{\text{3}}}} \right)^r}{\left( {\frac{{{\gamma _{th}}{P_c}{b_{\text{3}}}}}{{{P_r}}} + 1} \right)^{ - \left( {m + r} \right)}},
\end{align}
\emph{where we have ${b_{\text{1}}} = \frac{{{P_{BS}}{G_r}{C_r}{\gamma ^2}{\beta_{semi}^2}{B^2}\sigma _\tau ^2}}{{{G_c}{C_c}{{\left( {{d_r}} \right)}^{{\alpha _r} - {\alpha _c}}}}}$, ${b_{\text{2}}}=\frac{{{\sigma ^2}}}{{{G_c}{C_c}{{\left( {{d_r}} \right)}^{ - {\alpha _c}}}}}$, and ${b_{\text{3}}}=\frac{{{{\left( {{d_c}} \right)}^{ - {\alpha _c}}}}}{{{{\left( {{d_r}} \right)}^{ - {\alpha _c}}}}}$.}

Sketch of Proof:
By the accurate series expansion for the lower incomplete Gamma function and the binomial theorem, the OP expression is formulated as:
\begin{align}
 & \mathbb{P}_r^{II} = 1 - \sum\limits_{p = 0}^{m - 1} {\frac{1}{{p!}}} {\left( {\frac{{m{\gamma _{th}}}}{{{P_r}}}} \right)^p}\exp\left( { - \frac{{m{\gamma _{th}}\left( {{b_{\text{1}}}{\text{ + }}{b_{\text{2}}}} \right)}}{{{P_r}}}} \right)\notag\\
  &\hspace*{0.1cm}\times \sum\limits_{r = 0}^p {C_p^r} {\left( {{b_{\text{1}}}{\text{ + }}{b_{\text{2}}}} \right)^{p - r}}{\left( {{P_c}{b_{\text{3}}}} \right)^r} \notag\\
  &\hspace*{0.1cm}\times  \underbrace {\int_0^\infty  {{x^r}\exp \left( { - \frac{{m{\gamma _{th}}{P_c}{b_{\text{3}}}}}{{{P_r}}}x} \right)} {f_{{{\left| {{h_c}} \right|}^2}}}\left( x \right)dx}_{{I_2}} .
\end{align}

Furthermore, according to Eq. [2.3.3.1] in \cite{table}, we obtain the final analytical result. Additionally, the detailed proof is similar to that of \textbf{Theorem \ref{OP_I_c_f}}.
\end{theorem}

\begin{corollary}\label{OP_II_r_r}
\emph{We define a parameter of ${\Lambda _3} = \frac{{m\left( {{b_{\text{1}}}{\text{  +  }}{b_{\text{2}}}} \right)}}{{{P_r}}}$. When we consider the NOMA-based Semi-ISaC network in Scenario-II, the ergodic rate of the communication signal for the radar target is derived as:}
\begin{align}
 & R_r^{er,II} = \frac{1}{{\ln 2}}\sum\limits_{p = 0}^{m - 1} \sum\limits_{r = 0}^p \sum\limits_{k = 0}^\infty  \binom{m + r + k - 1}{k}\notag\\
 &\hspace*{0.1cm}\times  \frac{{C_p^r\Gamma \left( {m + r} \right)\Gamma \left( {p + k + 1} \right)}}{{\Gamma \left( m \right){m^r}p!{{\left( {{b_{\text{1}}}{\text{  +  }}{b_{\text{2}}}} \right)}^r}\Lambda _3^{k + 1}}}   \notag \\
  &\hspace*{0.1cm}\times {\left( {{P_c}{b_{\text{3}}}} \right)^r}{\left( { - \frac{{{P_c}{b_{\text{3}}}}}{{{P_r}}}} \right)^k}\exp \left( {{\Lambda _3}} \right){E_{p + k + 1}}\left( {{\Lambda _3}} \right) .
\end{align}

Sketch of Proof:
The proof is similar to that of \textbf{Corollary \ref{OP_I_c_r}}.
\end{corollary}

\begin{theorem}\label{OP_II_c_f}
\emph{Recall that we consider the NOMA-based Semi-ISaC network in Scenario-II. For the communication signal of the communication transmitter, the OP expression is formulated as:}
\begin{align}
&\mathbb{P}_c^{II} = 1 - \sum\limits_{p = 0}^{m - 1} {\frac{1}{{p!}}} {\left( {\frac{{m{\gamma _{SIC}}}}{{{P_r}}}} \right)^p}\exp\left( { - \frac{{m{\gamma _{SIC}}\left( {{b_{\text{1}}}{\text{ + }}{b_{\text{2}}}} \right)}}{{{P_r}}}} \right)\notag\\
&\hspace*{0.1cm}\times \sum\limits_{r = 0}^p {C_p^r} {\left( {{b_{\text{1}}}{\text{ + }}{b_{\text{2}}}} \right)^{p - r}}{\left( {{P_c}{b_{\text{3}}}} \right)^r}{I_3},
\end{align}
\emph{where ${I_3}$ is given by}
\begin{align}
&{I_3} = \frac{1}{{\Gamma \left( m \right){m^r}}}{\left( {\frac{{{\gamma _{SIC}}}}{{{P_r}}}{P_c}{b_{3}} + 1} \right)^{ - \left( {m + r} \right)}}\notag\\
&\hspace*{0.1cm}\times \Gamma \left( {m + r,\frac{{{\gamma _{th}}m}}{{{P_c}}}\left( {{a_1} + {a_2}} \right)\left( {\frac{{{\gamma _{SIC}}}}{{{P_r}}}{P_c}{b_{3}} + 1} \right)} \right).
\end{align}

Sketch of Proof:
By exploiting the series expansion of the incomplete Gamma function, the binomial theorem, and some equation manipulations, the OP of the communication transmitter in Scenario-II is
\begin{align}\label{eq35}
  &\mathbb{P}_c^{II} = 1 - \sum\limits_{p = 0}^{m - 1} {\frac{1}{{p!}}} {\left( {\frac{{m{\gamma _{SIC}}}}{{{P_r}}}} \right)^p}\exp\left( { - \frac{{m{\gamma _{SIC}}\left( {{b_{\text{1}}}{\text{ + }}{b_{\text{2}}}} \right)}}{{{P_r}}}} \right)\notag\\
  &\hspace*{0.1cm}\times \sum\limits_{r = 0}^p {C_p^r} {\left( {{b_{\text{1}}}{\text{ + }}{b_{\text{2}}}} \right)^{p - r}}{\left( {{P_c}{b_{\text{3}}}} \right)^r} \notag \\
&\hspace*{0.1cm}\times  \underbrace {\int_{\frac{{{\gamma _{th}}}}{{{P_c}}}\left( {{a_1} + {a_2}} \right)}^\infty  {{x^r}\exp \left( { - \frac{{m{\gamma _{SIC}}}}{{{P_r}}}{P_c}{b_{\text{3}}}x} \right)} {f_{{{\left| {{h_c}} \right|}^2}}}\left( x \right)dx}_{{I_3}}.
\end{align}

Then, we can derive the final OP expression by substituting Eq. [2.3.6.6] from \cite{table} into the expression as Eq. \eqref{eq35}. The detailed proof is similar to that of \textbf{Theorem \ref{OP_I_r_f}}.
\end{theorem}

\begin{corollary}\label{OP_II_c_r}
\emph{For Scenario-II, the ergodic rate expression of the communication signal for the communication transmitter is formulated as: }
\begin{align}
  &R_c^{er,II} = \frac{1}{{\ln 2}}\sum\limits_{p = 0}^{m - 1} {\sum\limits_{r = 0}^p {\frac{{C_p^r{{\left( {{P_c}{b_{\text{3}}}} \right)}^r}\Lambda _5^p\exp \left( { - {\Lambda _5}} \right)}}{{p!\Gamma \left( m \right){m^r}{{\left( {{b_{\text{1}}}{\text{  +  }}{b_{\text{2}}}} \right)}^r}}}} } \notag\\
  &\hspace*{0.1cm}\times {\left( {\frac{{{\gamma _{SIC}}}}{{{P_r}}}{P_c}{b_{\text{3}}} + 1} \right)^{ - \left( {m + r} \right)}} \sum\limits_{k = 0}^{m + r - 1} {\frac{{\left( {m + r - 1} \right)!}}{{k!{\Lambda _4}}}}\notag \\
  &\hspace*{0.1cm}\times  \exp \left( {{\Lambda _4}} \right)\Gamma \left( {k + 1} \right){E_{k + 1}}\left( {{\Lambda _4}} \right),
\end{align}
\emph{where we have ${\Lambda _4} = \frac{m}{{{P_c}}}\left( {{a_1} + {a_2}} \right)\left( {\frac{{{\gamma _{SIC}}}}{{{P_r}}}{P_c}{b_{\text{3}}} + 1} \right)$ and ${\Lambda _5} = \frac{{m{\gamma _{SIC}}\left( {{b_{\text{1}}}{\text{  +  }}{b_{\text{2}}}} \right)}}{{{P_r}}}$.}

Sketch of Proof: The proof is similar to that of \textbf{Corollary \ref{OP_I_c_r}}.
\end{corollary}

\subsection{Analytical Performance Evaluation for Radar Echoes}
As the radar echoes are simply left behind after the last SIC process, the definition of ergodic REIR is the same as the OMA-based Semi-ISaC scenario when the SIC processes are successful. That is, under a perfect SIC case, the derivations of the ergodic REIR in the NOMA-based Semi-ISaC scenario are the same as those in the OMA-based Semi-ISaC scenario in \textbf{Theorem \ref{OPEROMA}}. Hence, we will not repeat the derivations here. To gain further insights, the closed-form asymptotic expressions of the ergodic REIR are derived and evaluated in the next section.

\section{Asymptotic Performance Evaluation for NOMA-based Semi-ISaC}

In this section, we derive the asymptotic OP and the asymptotic ergodic REIR for further evaluating the performance of the NOMA-based Semi-ISaC system in the high-SNR region. We derive the diversity orders of the OP (for the communication signals) and present our insights in \textbf{Remark \ref{diversity1}-\ref{diversitym}}. Additionally, we derive the high-SNR slopes of the ergodic REIR (for the radar echoes) and summarize them in \textbf{Remark \ref{SNRSLOPE}-\ref{SNRSLOPE2}}.

\subsection{Asymptotic Outage Performance and Diversity Gains for Communication Signals}
Recall that we consider two scenarios, namely Scenario-I having a near communication transmitter and a far radar target and Scenario-II associated with a far communication transmitter and a near radar target.

\subsubsection{Diversity Evaluation in Scenario-I}
Based on \textbf{Theorem \ref{OP_I_c_f}} and \textbf{Theorem \ref{OP_I_r_f}}, we evaluate the performance in the high-SNR region. Explicitly, we exploit the asymptotic series of the lower incomplete Gamma function and retain only a single term as the following form of
\begin{align}\label{series}
\gamma \left( {a,b} \right) \approx \sum\limits_{n=0}^\infty  {\frac{{\left( { - 1} \right){b^{a + n}}}}{{n!\left( {a + n} \right)}} \approx \frac{{{b^a}}}{a}} .
\end{align}

Then, we substitute Eq. \eqref{series} into the results of \textbf{Theorem \ref{OP_I_c_f}} and \textbf{Theorem \ref{OP_I_r_f}}, and following some further manipulations, we arrive at the asymptotic OP expressions, which are presented in \textbf{Corollary \ref{OP_I_c_f_a}} and \textbf{Corollary \ref{OP_I_r_f_a}}.

\begin{corollary}\label{OP_I_c_f_a}
\emph{For the communication signal of the communication transmitter in Scenario-I, the asymptotic OP expression is }
\begin{align}
\mathbb{P}_{c,\infty }^I={\left( {\frac{{m{\gamma _{th}}}}{{{P_c}}}} \right)^m}\sum\limits_{r = 0}^m {C_m^r} {\left( {{a_1} + {a_2}} \right)^{m - r}}\frac{{{{\left( {{P_r}{a_3}} \right)}^r}\Gamma \left( {m + r} \right)}}{{\Gamma \left( {m + 1} \right)\Gamma \left( m \right){m^r}}}.
\end{align}

Sketch of Proof:
Upon Substituting Eq. \eqref{series} into the OP expression of \textbf{Theorem \ref{OP_I_c_f}}, we have
\begin{align}
&\mathbb{P}_{c,\infty }^I = {\left( {\frac{{m{\gamma _{th}}}}{{{P_c}}}} \right)^m}\sum\limits_{r = 0}^m {C_m^r} {\left( {{a_1} + {a_2}} \right)^{m - r}}\notag\\
&\hspace*{0.1cm}\times \frac{{{{\left( {{P_r}{a_3}} \right)}^r}}}{{\Gamma \left( {m + 1} \right)}} \int_0^\infty  {{x^r}} {f_{{{\left| {{h_r}} \right|}^2}}}\left( x \right)dx.
\end{align}

With the aid of the PDF of the Gamma distribution, we derive the integral as $\int_0^\infty  {{x^r}} {f_{{{\left| {{h_r}} \right|}^2}}}\left( x \right) $ $dx = \frac{{{m^m}}}{{\Gamma \left( m \right)}}\int_0^\infty  {{x^{m + r - 1}}} \exp \left( { - mx} \right)dx = \frac{{\Gamma \left( {m + r} \right)}}{{\Gamma \left( m \right){m^r}}}$. Then, after we substitute the integral into the OP expression, this proof is completed.
\end{corollary}

\begin{remark}\label{diversity1}
\emph{To evaluate the outage performance in the high-SNR region, we express the diversity order of the communication transmitter in Scenario-I as:}
\begin{align}
D_c^I =  - \mathop {\lim }\limits_{{P_c} \to \infty } \frac{{\log \left( {\mathbb{P}_{c,\infty }^I} \right)}}{{\log \left( {{P_c}} \right)}} = m,
\end{align}
\emph{which is proved by $\mathop {\lim }\limits_{x \to \infty } \frac{{\log \left[ {{{\left( {A/x} \right)}^m}} \right]}}{{\log \left( x \right)}} = m$ for a constant $A$ independent of the variable $x$. In the high-SNR region, the slope of the OP of the communication transmitter is $m$}
\end{remark}

\begin{corollary}\label{OP_I_r_f_a}
\emph{For the communication signal of the radar target in Scenario-I, the asymptotic OP is formulated as: }
\begin{align}
  &\mathbb{P}_{r,\infty }^I = {\text{ }}{F_{{{\left| {{h_r}} \right|}^2}}}\left( {\frac{{{\gamma _{th}}\left( {{a_4} + {a_5}} \right)}}{{{P_r}}}} \right) + {\left( {\frac{{m{\gamma _{SIC}}}}{{{P_c}}}} \right)^m}\notag\\
  &\hspace*{0.1cm}\times \sum\limits_{r = 0}^m {C_m^r} \frac{{{{\left( {{a_3}{P_r}} \right)}^r}\Gamma \left( {m + r,\frac{{m{\gamma _{th}}\left( {{a_4} + {a_5}} \right)}}{{{P_r}}}} \right)}}{{{{\left( {{a_1} + {a_2}} \right)}^{r - m}}\Gamma \left( {m + 1} \right)\Gamma \left( m \right){m^r}}}.
\end{align}

Sketch of Proof:
With the aid of \textbf{Theorem \ref{OP_I_r_f}}, we can formulate the OP expression as:
\begin{align}
&\mathbb{P}_{r,\infty }^I = {F_{{{\left| {{h_r}} \right|}^2}}}\left( {\frac{{{\gamma _{th}}\left( {{a_4} + {a_5}} \right)}}{{{P_r}}}} \right) + {\left( {\frac{{m{\gamma _{SIC}}}}{{{P_c}}}} \right)^m}\notag\\
&\hspace*{0.1cm}\times \sum\limits_{r = 0}^m {C_m^r} {\text{ }}\frac{{{{\left( {{a_3}{P_r}} \right)}^r}\int_{\frac{{{\gamma _{th}}\left( {{a_4} + {a_5}} \right)}}{{{P_r}}}}^\infty  {{x^r}} {f_{{{\left| {{h_r}} \right|}^2}}}\left( x \right)dx}}{{{{\left( {{a_1} + {a_2}} \right)}^{r - m}}\Gamma \left( {m + 1} \right)}},
\end{align}
and based on the PDF of the Gamma distribution and the integral $\frac{{{m^m}}}{{\Gamma \left( m \right)}}\int_{\frac{{{\gamma _{th}}\left( {{a_4} + {a_5}} \right)}}{{{P_r}}}}^\infty  {{x^{m + r - 1}}} $ $\times\exp \left( { - mx} \right)dx = \frac{1}{{\Gamma \left( m \right){m^r}}}\Gamma \left( {m + r,\frac{{m{\gamma _{th}}\left( {{a_4} + {a_5}} \right)}}{{{P_r}}}} \right)$, the final OP expression is derived.
\end{corollary}

\begin{remark}\label{diversity2}
\emph{For the high-SNR region in Scenario-I, based on the asymptotic expression of the radar target's communication signal, we derive the diversity order for the radar target as:}
\begin{align}
D_r^I =  - \mathop {\lim }\limits_{{P_c} \to \infty } \frac{{\log \left( {\mathbb{P}_{r,\infty }^I} \right)}}{{\log \left( {{P_c}} \right)}} = 0,
\end{align}
\emph{which is proved by $\mathop {\lim }\limits_{x \to \infty } \frac{{\log \left[ {{{\left( {A/x} \right)}^m} + B} \right]}}{{\log \left( x \right)}} = 0$ with the constants $A$ and $B$ that are independent of the variable $x$. The OP of the radar target catches the lower limit in the high-SNR region of Scenario-I.}
\end{remark}

\subsubsection{Diversity Evaluation in Scenario-II}
For Scenario-II of the NOMA-based Semi-ISaC network, based on the results of \textbf{Theorem \ref{OP_II_r_f}} and \textbf{Theorem \ref{OP_II_c_f}}, we are able to exploit the asymptotic series expansion of Eq. \eqref{series} for deriving the asymptotic OP. Thus, the asymptotic OP of the communication transmitter and radar target are given by \textbf{Corollary \ref{OP_II_c_f_a}} and \textbf{Corollary \ref{OP_II_r_f_a}}, respectively.

\begin{corollary}\label{OP_II_c_f_a}
\emph{In Scenario-II, the asymptotic OP expression of the communication transmitter is  }
\begin{align}
&\mathbb{P}_{c,\infty }^{II} = {F_{{{\left| {{h_c}} \right|}^2}}}\left( {\frac{{{\gamma _{th}}}}{{{P_c}}}\left( {{a_1} + {a_2}} \right)} \right){\text{  +  }}{\left( {\frac{{m{\gamma _{SIC}}}}{{{P_r}}}} \right)^m}\notag\\
&\hspace*{0.1cm}\times  \sum\limits_{r = 0}^m {C_m^r} \frac{{{{\left( {{P_c}{b_{\text{3}}}} \right)}^r}\Gamma \left( {m + r,\frac{{m{\gamma _{th}}}}{{{P_c}}}\left( {{a_1} + {a_2}} \right)} \right)}}{{{{\left( {{b_{\text{1}}}{\text{  +  }}{b_{\text{2}}}} \right)}^{r - m}}\Gamma \left( {m + 1} \right)\Gamma \left( m \right){m^r}}}.
\end{align}

Sketch of Proof:
The proof is similar to that of \textbf{Corollary \ref{OP_I_r_f_a}}.
\end{corollary}

\begin{remark}\label{diversity3}
\emph{In Scenario-II, we evaluate the outage performance in the high-SNR region by assuming that the transmit power of the radar target is infinity. Based on the \textbf{Corollary \ref{OP_II_c_f_a}}, the diversity order of the communication transmitter is expressed as:}
\begin{align}
D_c^{II} =  - \mathop {\lim }\limits_{{P_r} \to \infty } \frac{{\log \left( {\mathbb{P}_{c,\infty }^{II}} \right)}}{{\log \left( {{P_r}} \right)}} = 0,
\end{align}
indicating that the OP of the communication transmitter in Scneario-II has a lower bound.
\end{remark}

\begin{corollary}\label{OP_II_r_f_a}
\emph{In Scenario-II, the asymptotic OP expression of the radar target is derived as:  }
\begin{align}
\mathbb{P}_{r,\infty }^{II} {=} {\left( {\frac{{m{\gamma _{th}}}}{{{P_r}}}} \right)^m}\sum\limits_{r = 0}^m {C_m^r} {\left( {{b_{\text{1}}}{\text{ + }}{b_{\text{2}}}} \right)^{m - r}}\frac{{{{\left( {{P_c}{b_{\text{3}}}} \right)}^r}\Gamma \left( {m + r} \right)}}{{\Gamma \left( {m + 1} \right)\Gamma \left( m \right){m^r}}}.
\end{align}

Sketch of Proof:
The proof is similar to that of \textbf{Corollary \ref{OP_I_c_f_a}}.
\end{corollary}

\begin{remark}\label{diversity4}
\emph{Under the same assumptions as in \textbf{Remark \ref{diversity3}}, we exploit the asymptotic expressions yielding the diversity order of the communication signal of the radar target in Scenario-II as:}
\begin{align}
D_r^{II} =  - \mathop {\lim }\limits_{{P_r} \to \infty } \frac{{\log \left( {\mathbb{P}_{r,\infty }^{II}} \right)}}{{\log \left( {{P_r}} \right)}} = m,
\end{align}
showing that the communication signal is directly influenced by the LoS component $m$.
\end{remark}

\begin{remark} \label{diversitym}
For Nakagami-m fading channels, we conclude that with a strong LoS component (large $m$), we have high diversity orders, yielding a Neal-Gaussian performance reminiscent of an asymptotically infinite diversity order.
\end{remark}

\subsection{Asymptotic Ergodic REIR and High-SNR Slopes}

We exploit the asymptotic expansions of the lower incomplete Gamma function and the generalized exponential integral to derive the asymptotic ergodic REIR for the radar target, expressed as
$\gamma \left( {m,t} \right) = \left( {m - 1} \right)! - \exp \left( { - t} \right)\sum\limits_{k = 0}^{m - 1} {\frac{{\left( {m - 1} \right)!}}{{k!}}{t^k}} $, ${E_n}\left( z \right) \approx \frac{{{{\left( { - z} \right)}^{n - 1}}}}{{\left( {n - 1} \right)!}}\left( {\psi \left( n \right)} \right. -$ $\left. {  \ln \left( z \right)} \right) - \sum\limits_{k = 0{\text{\& }}k \ne n - 1} {\frac{{{{\left( { - z} \right)}^k}}}{{k!\left( {1 - n + k} \right)}}} $ for $n>1$, and ${E_1}\left( z \right) \approx  - {C_\gamma } - \ln \left( z \right) + z$, where ${C_\gamma }$ is the Euler constant and ${\psi \left( n \right)}$ is the Psi function. The asymptotic expression of the ergodic REIR for radar target is given in \textbf{Corollary \ref{AER1}}. We further evaluate the high-SNR slope in \textbf{Remark \ref{SNRSLOPE}}.

\begin{corollary}\label{AER1}
\emph{Upon assuming that $m$ is an integer denoted as $m \in \mathbb{Z}$, we derive the closed-form asymptotic expression of the ergodic REIR as:}
\begin{align}\label{Cfinal}
R_{est}^\infty =\frac{{\delta {m^m}}}{{2T\ln \left( 2 \right)\Gamma \left( m \right)}}{I_4} + \sum\limits_{k = {\text{1}}}^{m - 1} {\frac{{\delta {m^m}}}{{2T\ln \left( 2 \right)\Gamma \left( m \right)}}} {I_{5}},
\end{align}
\emph{where $I_4$ and $I_{5}$ are formulated as:}
\begin{align}
  &{I_4} {=} \frac{{d_t^{{\alpha _r}}\Gamma \left( {m {-} 1} \right)}}{{{\Xi _{r,1}}{m^{m {-} 2}}}} {-} \frac{{\Gamma \left( m \right)}}{{{m^m}}}\left( {\log \left( {\frac{{{m^2}d_t^{{\alpha _r}}}}{{{\Xi _{r,1}}}}} \right) {-} {\psi^{\left( 0 \right)}}\left( m \right) {-} {C_\gamma }} \right) \notag \\
   &\hspace*{0.1cm}+ {\left( {\frac{{md_t^{{\alpha _r}}}}{{{\Xi _{r,1}}}}} \right)^2}\frac{{\Gamma \left( {m - 2} \right)}}{{{m^{m - 2}}}} - \frac{{md_t^{{\alpha _r}}}}{{{\Xi _{r,1}}}}\frac{{\Gamma \left( {m - 1} \right)}}{{{m^{m - 1}}}}\notag\\
  & \hspace*{0.1cm} \times \left( {\log \left( {\frac{{{m^2}d_t^{{\alpha _r}}}}{{{\Xi _{r,1}}}}} \right) - {\psi ^{\left( 0 \right)}}\left( {m - 1} \right) - {C_\gamma }} \right)  ,\\
  &{I_{5}} {=} {\left( { - \frac{{d_t^{{\alpha _r}}}}{{{\Xi _{r,1}}}}} \right)^k}\frac{{\psi \left( {k + 1} \right)\Gamma \left( {m - k} \right)}}{{k!{m^m}}}\notag\\
   &\hspace*{0.1cm} + \sum\limits_{q = 0\& q \ne k}^{m - 1} {\frac{{\Gamma \left( {m - q} \right){{\left( { - \frac{{md_t^{{\alpha _r}}}}{{{\Xi _{r,1}}}}} \right)}^q}}}{{q!\left( {q - k} \right){m^{m - q}}}}} {-} \frac{{\Gamma \left( {m - k} \right)}}{{{m^{m - k}}k!}}{\left( { {-} \frac{{md_t^{{\alpha _r}}}}{{{\Xi _{r,1}}}}} \right)^k} \notag\\
   &\hspace*{0.1cm}\times  \left( {\ln \left( {\frac{{{m^2}d_t^{{\alpha _r}}}}{{{\Xi _{r,1}}}}} \right) - {\psi ^{\left( 0 \right)}}\left( {m - k} \right)} \right) .
\end{align}

Sketch of Proof:
See Appendix~C.
\end{corollary}

With the aid of the derived asymptotic expressions, we then evaluate the high-SNR slope of the radar target. Conditioned on $P_{BS} \to \infty $, the high-SNR slope is defined as $S = \mathop {\lim }\limits_{{P_{BS}} \to \infty } \frac{{R_{est,\infty }^{low}\left( {{P_{BS}}} \right)}}{{\ln \left( {{P_{BS}}} \right)}}.$

\begin{remark}\label{SNRSLOPE}
Upon substituting the expression in \textbf{Corollary \ref{AER1}} into the high-SNR slope definition, the high-SNR slope is formulated as:
\begin{align}
S =& \mathop {\lim }\limits_{{P_{BS}} \to \infty } \frac{{\frac{{\delta {m^m}}}{{2T\ln \left( 2 \right)\Gamma \left( m \right)}}{I_7}}}{{\ln \left( {{P_{BS}}} \right)}} + \mathop {\lim }\limits_{{P_{BS}} \to \infty } \frac{{\frac{{\delta {m^m}}}{{2T\ln \left( 2 \right)\Gamma \left( m \right)}}{I_{8}}}}{{\ln \left( {{P_{BS}}} \right)}} \notag\\
=& \frac{\delta }{{2T\ln \left( 2 \right)}},
\end{align}
which is proved by exploiting equations $\mathop {\lim }\limits_{{P_{BS}} \to \infty } \frac{{{{\left( {\frac{A}{{{P_{BS}}}}} \right)}^k} - B\ln \left( {\frac{C}{{{P_{BS}}}}} \right) + D}}{{\ln \left( {{P_{BS}}} \right)}} = B$ and $\mathop {\lim }\limits_{{P_{BS}} \to \infty } \frac{{{{\left( {\frac{A}{{{P_{BS}}}}} \right)}^k}}}{{\ln \left( {{P_{BS}}} \right)}} - \frac{{{{\left( {\frac{B}{{{P_{BS}}}}} \right)}^k}\ln \left( {\frac{C}{{{P_{BS}}}}} \right)}}{{\ln \left( {{P_{BS}}} \right)}}$ $ + \frac{D}{{\ln \left( {{P_{BS}}} \right)}} = 0$, where $A$, $B$, $C$, and $D$ are constants that are independent of the variable $P_{BS}$.
\end{remark}

\begin{remark}\label{SNRSLOPE2}
The high-SNR slope is only influenced by the radar's duty cycle $\delta$ and the pulse duration $T$. Additionally, the high-SNR slope is proportional to $\delta/T$.
\end{remark}

\section{Numerical Results}

Our numerical analysis is presented in this section, where the parameters are set as: the distance of the near-user is 800 meters and that of the far-user is 1300 meters, the bandwidth is $B=10$ MHz, the noise power is $\sigma^2= k_b \beta_{semi}B T_{temp}$ with $T_{temp}=724$ K, the target rates of the wireless communications are $\hat{R}_{OMA}= \hat{R}_{NOMA}= 1$ given the thresholds $\gamma_{th}=2^{\hat{R}_{NOMA}}-1$ for NOMA and $\gamma_{th}^{OMA}=2^{\hat{R}_{OMA}}-1$ for OMA, the threshold for SIC is $\gamma_{SIC}=0.4$, the carrier frequency is $f_c = 10^9$ Hz, the speed of light is $c = 3\times10^8$ m/s, the radar target's cross section is $\sigma_{RCS}=0.1$, the pulse duration is $T=1$ $\mu$s, the path loss exponents are $\alpha_r=4.5$ and $\alpha_c=2.5$, the radar's duty cycle is $\delta = 0.01$, and the Nagakami coefficient is $m=3$. We define the average received SINR as $\mathbb{E}\left[P_j\left|h_j\right|^2 C_j d_j^{-\alpha_j}/\sigma^2\right]$ (dB) where we exploit the subscript of $j\in \{c,r\}$ for representing the communication channels or the radar detection channels. For the SIC settings, we consider perfect SIC associated with the parameters ${{\varsigma _c}}=0$ and ${{\varsigma _r}}=0$. We will investigate the realistic imperfect SIC cases on our future research.

\subsection{From OMA-based Semi-ISaC to NOMA-based Semi-ISaC}

\begin{figure}[htbp]
\centering
\subfigure[]
{
	\begin{minipage}{7cm}
	\centering
	\includegraphics[width= 3.2in]{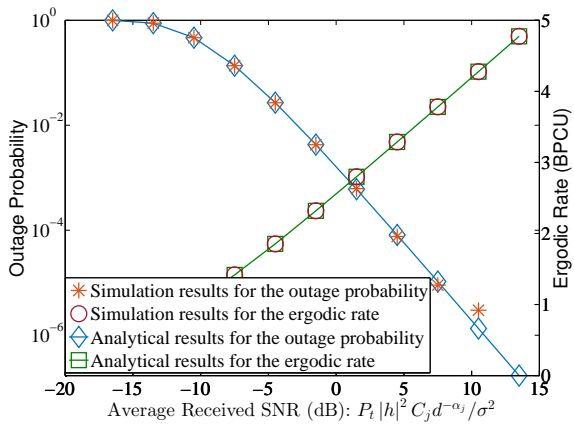}
    \label{figure1}
	\end{minipage}
}
\subfigure[]
{
	\begin{minipage}{7cm}
	\centering
	\includegraphics[width= 3.2in]{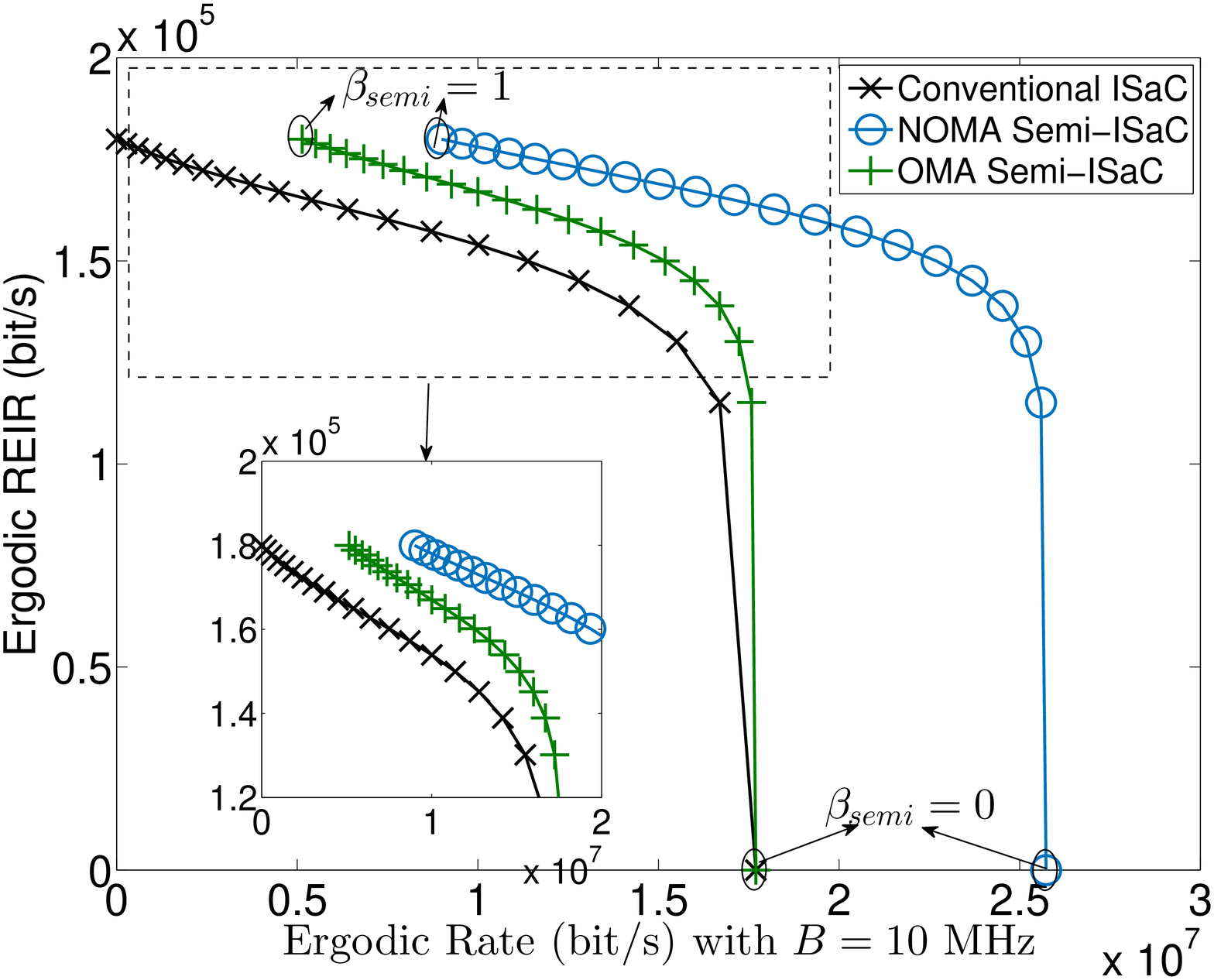}
    \label{figure2}
	\end{minipage}
}
\caption{From OMA to NOMA: (a) The verification of the OP and the ergodic rate for the OMA-based Semi-ISaC system based on \textbf{Theorem \ref{OPEROMA}-\ref{EREIR1}}; (b) A comparison among the conventional (FD) ISaC, OMA-based Semi-ISaC, and NOMA-based Semi-ISaC and the interplay between the radar target (ergodic REIR) and the communication transmitter (ergodic rate) with various $\beta_{semi} \in [0,1]$.}
\end{figure}

\begin{figure}[htbp]
\centering
\subfigure[]
{
	\begin{minipage}{7cm}
	\centering
	\includegraphics[width= 3.16in]{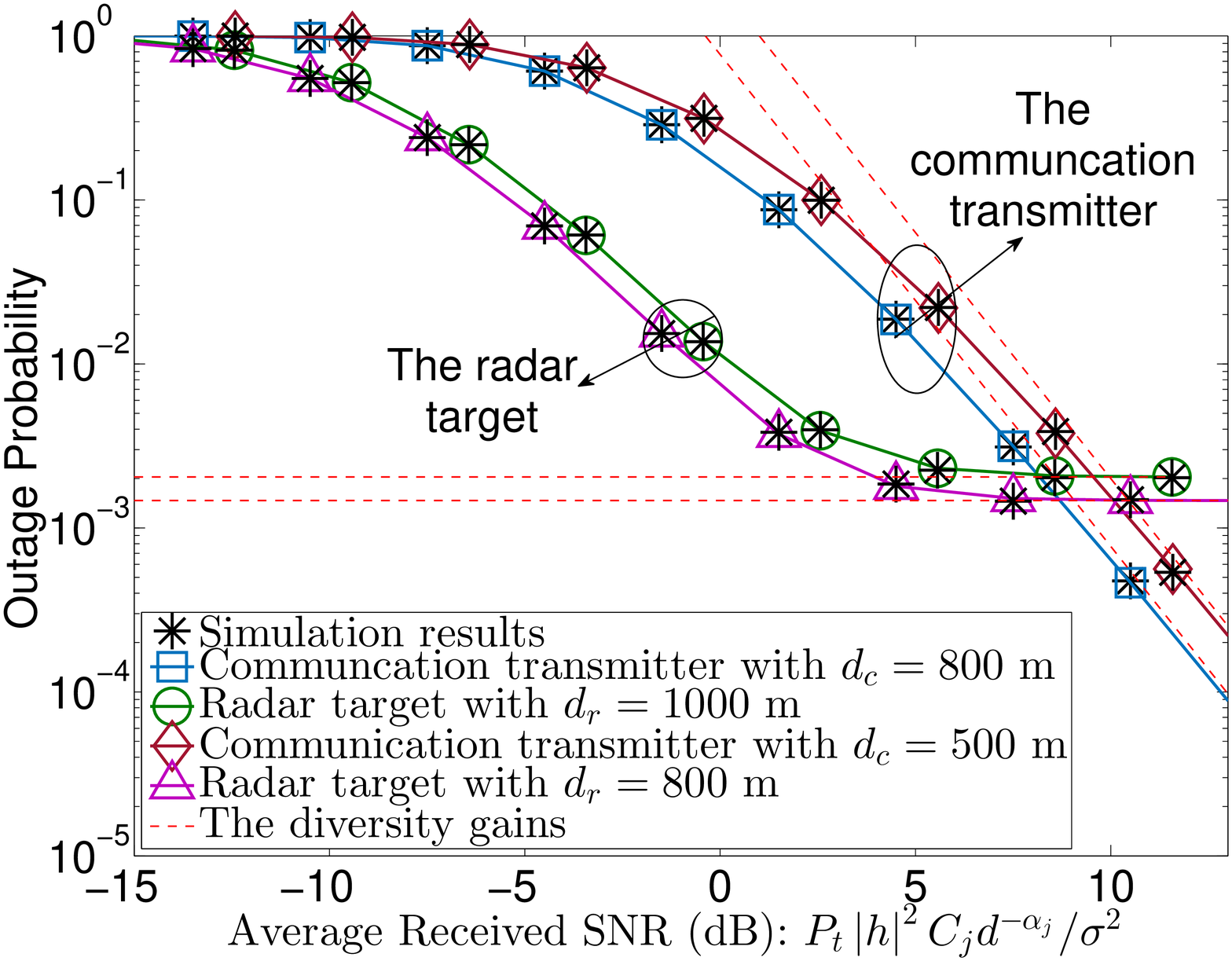}
    \label{figure3}
	\end{minipage}
}
\subfigure[]
{
	\begin{minipage}{7cm}
	\centering
	\includegraphics[width= 3.16 in]{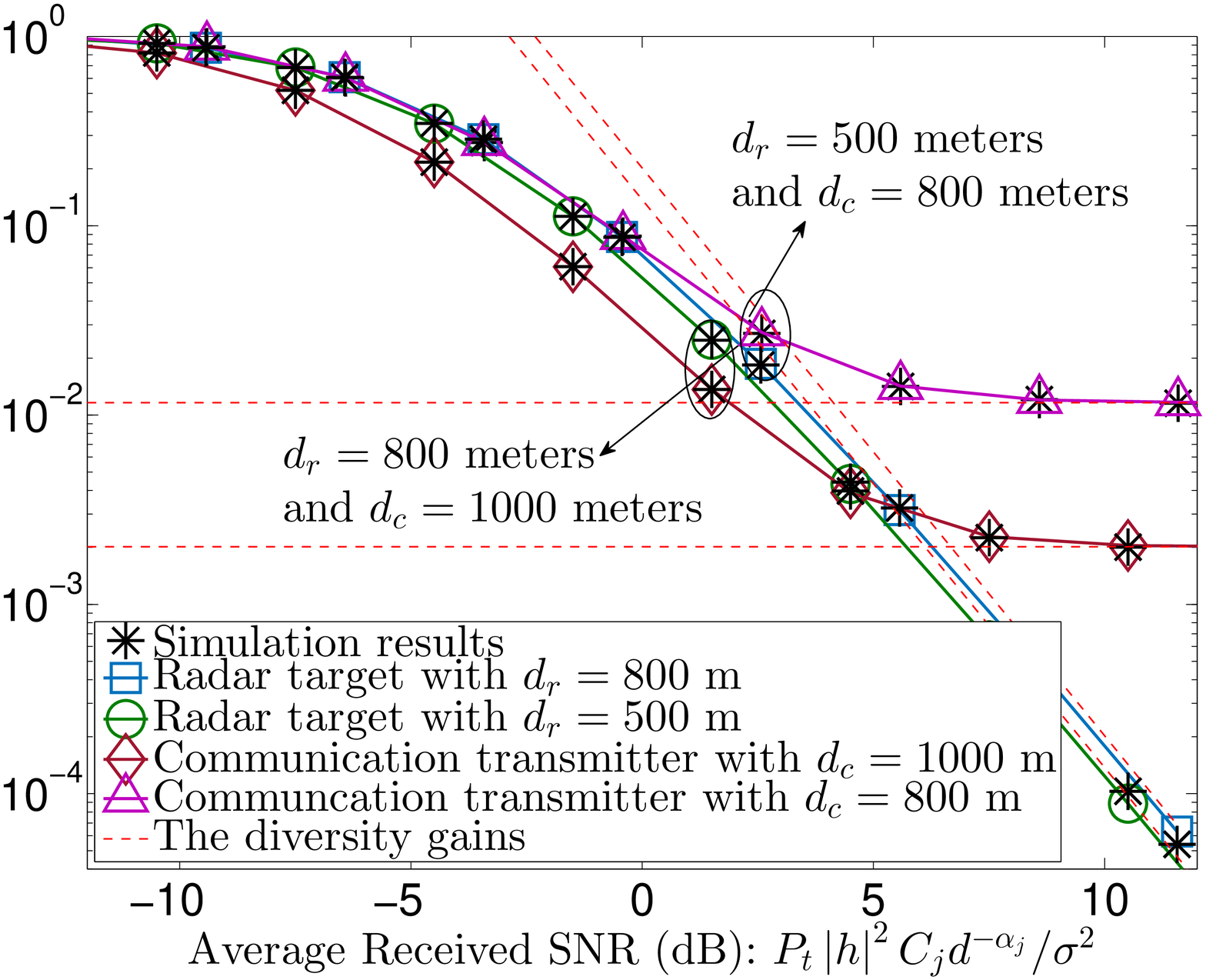}
    \label{figure4}
	\end{minipage}
}
\caption{Outage performance: (a) OP versus the received SNR of the communication transmitter in Scenario-I; (b) OP versus the received SNR of the radar target in Scenario-II. The analytical results are based on \textbf{Theorem \ref{OP_I_c_f}-\ref{OP_II_c_f}}, \textbf{Corollary \ref{OP_I_c_r}-\ref{OP_II_r_f_a}}, and \textbf{Remark \ref{diversity1}-\ref{diversitym}}.}
\end{figure}

\begin{figure}[htbp]
\centering
\subfigure[]
{
	\begin{minipage}{7cm}
	\centering
	\includegraphics[width= 3.2in]{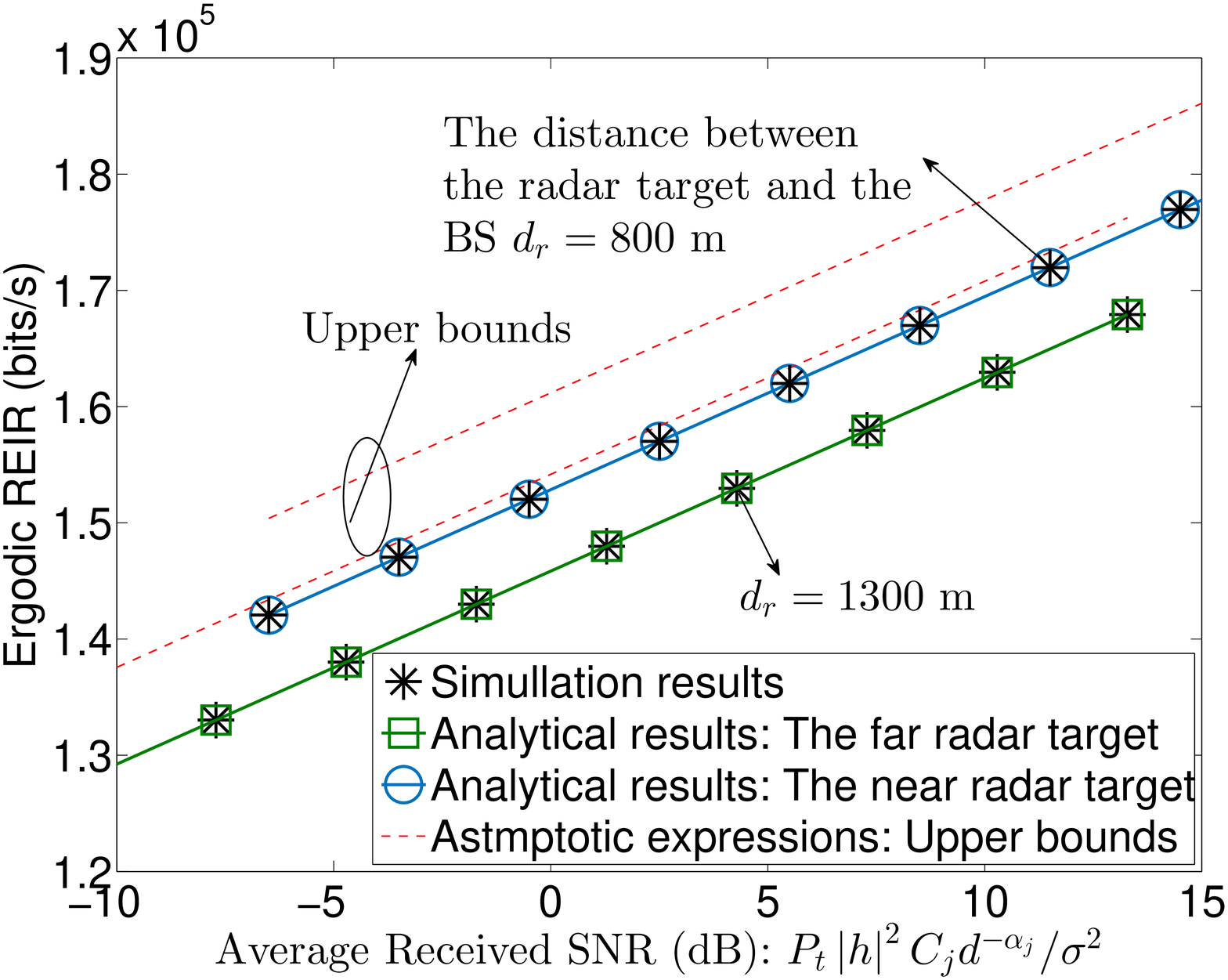}
    \label{figure5}
	\end{minipage}
}
\subfigure[]
{
	\begin{minipage}{7cm}
	\centering
	\includegraphics[width= 3.2 in, height = 2.2in]{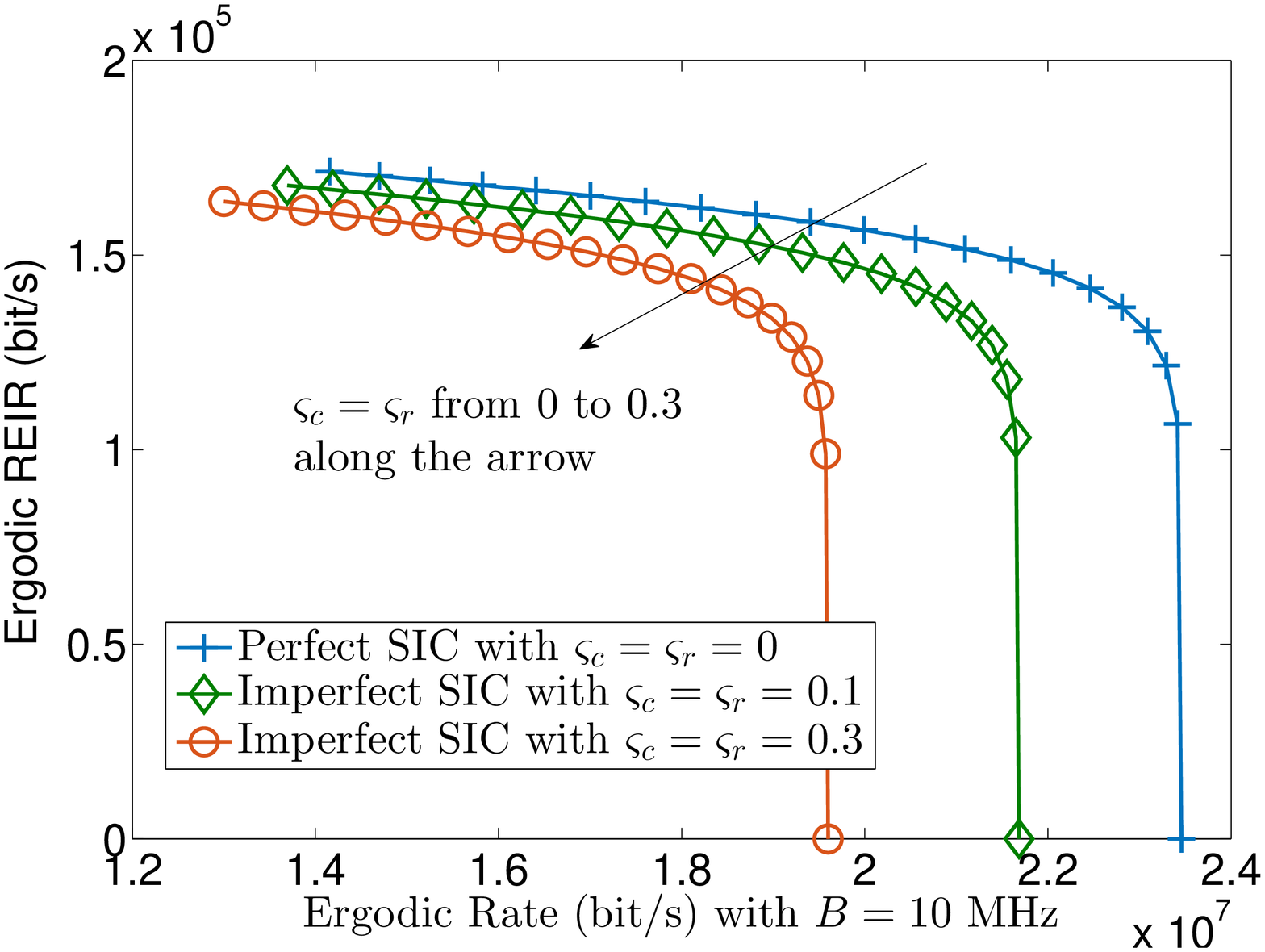}
    \label{figure6}
	\end{minipage}
}
\caption{The ergodic REIR: (a) The ergodic REIR versus the received SNR of the BS with various distance $d_r = [800,1300]$ meters; (b) The comparison between NOMA-based Semi-ISaC networks with perfect SIC and imperfect SIC. }
\end{figure}

In Fig. \ref{figure1}, we validate the OP and the ergodic rate (with the unit as \emph{bit per cell use}, denoted as BPCU) expressions of the communication signals versus the received power level ($d=800$ meters) under the OMA-based Semi-ISaC scenario ($P_t = 20$ dBm and $P_{BS}=[5,30]$ dBm). It can be seen that the analytical results are closely matched by the simulation results and there is no lower or higher limit of the metrics. This is because the interference arising from the radar signals is fixed and it is not increased when the OMA user's transmit power is increased. In Fig. \ref{figure2}, we investigate the performance interplay between the radar detection and wireless communications for $\beta_{semi} \in [0,1]$ when we set $\alpha_{semi}=0$ and $\epsilon_{semi} = 1- \beta_{semi} $. We then set the transmit power to 10 dBm for both the users and the BS. By comparing the performance among conventional (FD) ISaC, OMA-based Semi-ISaC, and NOMA-based Semi-ISaC networks, we observe that Semi-ISaC has better channel capacity than the conventional FD ISaC. The reason is that for Semi-ISaC, the radar and communication signals share the same resource blocks in ISaC bandwidth with the aid of the SIC to obtain better BE than that of the conventional FD ISaC. We also concluded that the NOMA-based Semi-ISaC scenario has a higher capacity than the OMA-based Semi-ISaC because the BE is further enhanced by the NOMA technique to share the resources by multiple communication users. Additionally, the ergodic REIR (for radar echoes) is zero when we have $\beta_{semi}=0$ because all the bandwidth is used for wireless communication and no bandwidth is set aside for radar detection. Thus we only have non-zero ergodic rate with zero ergodic RIER. Upon considering $\beta_{semi}=1$, we have the highest ergodic REIR, while the ergodic rate (for communication signals) cannot be reduced to zero. This represents the ISaC scenario (not Semi-ISaC), where the total bandwidth is utilized both for radar detection and for wireless communication.

\subsection{Outage Probability for Communication Signals in NOMA-based Semi-ISaC}

We validate the analytical and asymptotic OP expressions of NOMA users in Fig. \ref{figure3} and Fig. \ref{figure4} under $P_{BS} = 10$ dBm. Explicitly, in Fig. \ref{figure3}, a close communication transmitter ($d_c = \{500,800\}$ meters) and a distant radar target ($d_r = \{800,1000\}$ meters) are considered as Scenario-I ($P_c = [5,35]$ dBm and $P_r=20$ dBm). By contrast, in Fig. \ref{figure4}, our Scenario-II of a distant communication transmitter ($d_c = \{800,1000\}$ meters) and a close radar target ($d_r = \{500,800\}$ meters) is considered ($P_r = [5,35]$ dBm and $P_c=20$ dBm). We first observe that the simulation results closely match the analytical results and the diversity analysis matches the OP performance in the high-SNR region. Additionally, a conclusion for both scenarios is that upon increasing the near-user's received SNR, the OP of both users will be reduced while the far-user has an OP floor. The reason is that increasing the near-user's signal strength can enhance its received SNR directly. Additionally, the interference of the near-user is not increased, hence resulting in the reduced OP. By contrast, it beneficially improves the far-user's error rate of SIC by enhancing the near-user's received SNR, which only improves the OP of the distant user to a lower limit. But once the SIC process becomes perfect, the lower OP limit is reached.

\subsection{Ergodic REIR for Radar Echoes in NOMA-based Semi-ISaC}

In Fig. \ref{figure5}, the ergodic REIR of NOMA users is quantified. The analytical results fit the simulation results well and the asymptotic results represent the upper bound of the simulation results. Based on \textbf{Remark \ref{SNRSLOPE}}, the high-SNR slope is influenced by the ratio of the radar's duty cycle to pulse duration ($\delta/T$). In Fig. \ref{figure6} under the settings of Fig. \ref{figure2}, we compare the performance of NOMA-based Semi-ISaC networks both with and without perfect SIC. The figure indicates that the perfect SIC scenario represents the upper bounds compared to the imperfect SIC scenarios. With the interference left from the SIC process, both the ergodic rate of communication signals and the ergodic REIR of radar sensing are lower than those metrics with perfect SIC. This is because the SIC process will have errors in practical scenarios, hence the signals might become erroneously detected.

\section{Conclusions}

We have proposed the Semi-ISAC concept, where the total bandwidth is split as the radar-echo-only bandwidth, the communication-only bandwidth, and the ISaC bandwidth. We have evolved our novel Semi-ISaC concept from OMA to NOMA. We have then characterized the novel ergodic REIR metric for quantifying the average radar estimation rate. We have derived the OP and the ergodic rate for the communication signals and the ergodic REIR for the radar echo in the OMA-based Semi-ISaC scenario. In the NOMA-based Semi-ISaC scenario, we have derived the analytical expressions of the OP and the ergodic rate for communication signals. We have also derived the asymptotic OP along with the diversity gains attained for communication signals and the analytical expressions of the ergodic REIR for the radar echo, followed by the asymptotic ergodic REIR along with the high-SNR slopes. Our analysis has confirmed that: 1) The channel capacity of the conventional ISaC is lower than that of Semi-ISaC. 2) NOMA-based Semi-ISaC has better capacity than OMA-based Semi-ISsC; 3) The diversity gain of the communication signal is determined by the power of the line-of-sight component $m$; and 4) We can strike a flexible trad-off by balancing the radar and communication signals upon jointly controlling the transmit power of the BS, the radar's duty cycle, and the pulse duration. We will consider how to design the algorithms for predicting the radar echoes as our future research. We will also extend the perfect SIC case to the imperfect SIC scenario in our future research.

\section*{Appendix~A: Proof of Theorem~\ref{OP_I_c_f}} \label{Appendix:A}
\renewcommand{\theequation}{A.\arabic{equation}}
\setcounter{equation}{0}

For deriving the closed-form OP expressions for the communication transmitter, the probability expression should be manipulated as follows:
\begin{align}
&\mathbb{P}_c^I = \Pr \left\{ {\gamma _c^I < {\gamma _{th}}} \right\}\notag\\
&{=}\Pr \left\{ {{{\left| {{h_c}} \right|}^2} {<} \frac{{{\gamma _{th}}{P_r}{{\left| {{h_r}} \right|}^2}{{\left( {{d_r}} \right)}^{ {-} {\alpha _c}}}}}{{{P_c}{{\left( {{d_c}} \right)}^{ {-} {\alpha _c}}}}} {+} \frac{{{\gamma _{th}}\mathbb{E}\left[ {{I_R}} \right]\left( {{d_r}} \right) {+} {\gamma _{th}}{\sigma ^2}}}{{{P_c}{G_c}{C_c}{{\left( {{d_c}} \right)}^{ {-} {\alpha _c}}}}}} \right\}.
\end{align}

Upon substituting the expectation of interference in \textbf{Lemma \ref{AveragedInterference}} and rewriting the probability equation in form of integrals, we present the OP expression by exploiting the PDF and CDF of the Nakagami-\emph{m} fading channels as shown at the top of the next page, denoted as Eq. \eqref{A1}.
\begin{figure*}
\begin{align}\label{A1}
\mathbb{P}_c^I = \int_0^\infty  {\frac{1}{{\Gamma \left( m \right)}}\gamma \left( {m,m\left( {\frac{{{\gamma _{th}}{P_r}x{{\left( {{d_r}} \right)}^{ - {\alpha _c}}}}}{{{P_c}{{\left( {{d_c}} \right)}^{ - {\alpha _c}}}}} + \frac{{{\gamma _{th}}\mathbb{E}\left[ {{I_R}} \right]\left( {{d_r}} \right) + {\gamma _{th}}{\sigma ^2}}}{{{P_c}{G_c}{C_c}{{\left( {{d_c}} \right)}^{ - {\alpha _c}}}}}} \right)} \right)} {f_{{{\left| {{h_r}} \right|}^2}}}\left( x \right)dx.
\end{align}
\hrulefill
\end{figure*}

Since the CDF of the Nakagami-\emph{m} fading channel (in power domain) is a lower incomplete Gamma function, the accurate series expansion of the incomplete Gamma function may be exploited for reducing the complexity of derivation, which is expressed as:
 \begin{align}
\gamma \left( {a,b} \right) &{=} \Gamma \left( a \right) - \Gamma \left( {a,b} \right) {=} \Gamma \left( a \right) - \sum\limits_{p = 0}^{a - 1} {\frac{{\left( {a - 1} \right)!}}{{p!}}} \exp \left( { - b} \right){b^p},
\end{align}
where $\Gamma \left( {a,b} \right)$ is the upper incomplete Gamma function.

By substituting this equation into Eq. \eqref{A1}, we obtain the further streamlined expressions of
 \begin{align}\label{A2}
 & \mathbb{P}_c^I = {\text{1}} {-} \exp \left( { {-} \frac{{m{\gamma _{th}}}}{{{P_c}}}\left( {{a_1} + {a_2}} \right)} \right)\sum\limits_{p = 0}^{m - 1} \int_0^\infty {f_{{{\left| {{h_r}} \right|}^2}}}\left( x \right) \notag\\
  &\times{\frac{{{{\left( {m{\gamma _{th}}} \right)}^p}}}{{p!}}} {\exp \left( { {-} \frac{{m{\gamma _{th}}{P_r}{a_3}}}{{{P_c}}}x} \right)} {\left( {\frac{{{P_r}{a_3}x}}{{{P_c}}} {+} \frac{{{a_1} + {a_2}}}{{{P_c}}}} \right)^p}dx .
\end{align}

The former expression Eq. \eqref{A2} is then formulated with the aid of the Binomial theorem as:
\begin{align}
  &\mathbb{P}_c^I = {\text{1}} - \exp \left( { - \frac{{m{\gamma _{th}}}}{{{P_c}}}\left( {{a_1} + {a_2}} \right)} \right)\notag\\
  & \hspace*{0.1cm} \times \sum\limits_{p = 0}^{m - 1} {\frac{{{{\left( {m{\gamma _{th}}} \right)}^p}}}{{p!}}} \sum\limits_{r = 0}^p {C_p^r\frac{{{{\left( {{a_1} + {a_2}} \right)}^r}{{\left( {{P_r}{a_3}} \right)}^{p - r}}}}{{P_c^p}}}  \notag \\
 & \hspace*{0.1cm} \times \int_0^\infty  {\exp \left( { - \frac{{m{\gamma _{th}}{a_3}{P_r}}}{{{P_c}}}x} \right)} {x^{p - r}}{f_{{{\left| {{h_r}} \right|}^2}}}\left( x \right)dx.
\end{align}

We now exploit Eq. [2.3.3.1] of \cite{table} to obtain Eq. \eqref{Afinal}. Then, the proof is completed.

\section*{Appendix~B: Proof of Theorem~\ref{OP_I_r_f}} \label{Appendix:B}
\renewcommand{\theequation}{B.\arabic{equation}}
\setcounter{equation}{0}

The OP for the radar target under the NOMA-based Semi-ISaC scenario is expressed as the top of the next page, denoted as Eq. \eqref{B1}.
\begin{figure*}
\begin{align} \label{B1}
\mathbb{P}_r^I = 1 - \Pr \left\{ {{{\left| {{h_c}} \right|}^2} > {\gamma _{SIC}}\frac{{{a_3}{P_r}{{\left| {{h_r}} \right|}^2} + {a_1} + {a_2}}}{{{P_c}}},{{\left| {{h_r}} \right|}^2} > \frac{{{\gamma _{th}}\left( {{a_4} + {a_5}} \right)}}{{{P_r}}}} \right\}.
\end{align}
\hrulefill
\end{figure*}

Substituting the CDF of the Nakagami-\emph{m} fading channels into Eq. \ref{B1}, the resultant probability expression can be further transformed to Eq. \eqref{B2} at the top of the next page.
\begin{figure*}
\begin{align}\label{B2}
\mathbb{P}_r^I = 1 - \int_{\frac{{{\gamma _{th}}\left( {{a_4} + {a_5}} \right)}}{{{P_r}}}}^\infty  {\left( {1 - \frac{1}{{\Gamma \left( m \right)}}\gamma \left( {m,\frac{{m{\gamma _{SIC}}}}{{{P_c}}}\left( {{a_3}{P_r}{{\left| {{h_r}} \right|}^2} + {a_1} + {a_2}} \right)} \right)} \right)} {f_{{{\left| {{h_r}} \right|}^2}}}\left( x \right)dx.
\end{align}
\hrulefill
\end{figure*}

By exploiting an accurate series expansion of the lower incomplete Gamma function, and then further manipulating the equations, the OP expression is derived as:
\begin{align}\label{B3}
 & \mathbb{P}_r^I = 1 - \sum\limits_{p = 0}^{m - 1} {\frac{1}{{p!}}{{\left( {\frac{{m{\gamma _{SIC}}}}{{{P_c}}}} \right)}^p}\exp \left( { - \frac{{m{\gamma _{SIC}}\left( {{a_1} + {a_2}} \right)}}{{{P_c}}}} \right)}\notag\\
  & \hspace*{0.1cm} \times \sum\limits_{r = 0}^p {C_p^r} {{\left( {{a_1} + {a_2}} \right)}^{p - r}}{{\left( {{a_3}{P_r}} \right)}^r}  \notag \\
  & \hspace*{0.1cm} \times \underbrace {\int_{\frac{{{\gamma _{th}}\left( {{a_4} + {a_5}} \right)}}{{{P_r}}}}^\infty  {\exp \left( { - \frac{{m{\gamma _{SIC}}{a_3}{P_r}x}}{{{P_c}}}} \right)} {x^r}{f_{{{\left| {{h_r}} \right|}^2}}}\left( x \right)dx}_{{I_1}}.
\end{align}

Then we can derive $I_1$ based on Eq. [2.3.6.6] of \cite{table}, yielding:
\begin{align}\label{B4}
 & {I_1} =\frac{1}{{\Gamma \left( m \right){m^r}}}{\left( {\frac{{{\gamma _{SIC}}{a_3}{P_r}}}{{{P_c}}} + 1} \right)^{ - (r + m)}}\notag\\
   & \hspace*{0.1cm} \times \Gamma \left( {r + m,\frac{{{\gamma _{th}}m\left( {{a_4} + {a_5}} \right)}}{{{P_r}}}\left( {\frac{{{\gamma _{SIC}}{a_3}{P_r}}}{{{P_c}}} + 1} \right)} \right).
\end{align}

Finally, upon substituting $I_1$ from Eq. \eqref{B4} into the OP expression of Eq. \eqref{B3}, we can obtain the closed-form expression Eq. \eqref{Bfinal}. This completes the proof.

\section*{Appendix~C: Proof of Corollary~\ref{AER1}} \label{Appendix:C}
\renewcommand{\theequation}{C.\arabic{equation}}
\setcounter{equation}{0}

We first express the ergodic REIR with the aid of the following integrals as:
\begin{align}
&{R_{est}} \approx \sum\limits_{k = 0}^{m - 1} {\frac{\delta }{{2T\ln \left( 2 \right)}}} \int_0^\infty  {{{\left( {\frac{{md_t^{{\alpha _r}}}}{{{\Xi _{r,1}}x}}} \right)}^k}}\notag\\
 & \hspace*{0.1cm} \times \int_0^\infty  {\frac{1}{{z + 1}}\exp \left( { - \frac{{md_t^{{\alpha _r}}}}{{{\Xi _{r,1}}x}}z} \right)\frac{{{z^k}}}{{k!}}dz} {f_{{{\left| {{h_{r,u}}} \right|}^2}}}\left( x \right)dx .
\end{align}

With the aid of Eq. [2.3.6.9] of \cite{table}, we have
\begin{align}
&{R_{est}} \approx \sum\limits_{k = 0}^{m - 1} {\frac{\delta }{{2T\ln \left( 2 \right)}}} \int_0^\infty  {{\left( {\frac{{md_t^{{\alpha _r}}}}{{{\Xi _{r,1}}x}}} \right)}^k}\notag\\
& \hspace*{0.1cm} \times \Psi \left( {k + 1,k + 1,\frac{{md_t^{{\alpha _r}}}}{{{\Xi _{r,1}}x}}} \right){f_{{{\left| {{h_{r,u}}} \right|}^2}}}\left( x \right)dx ,
\end{align}
and based on $\Psi \left( {a,a,z} \right) = {z^{1 - a}}\exp \left( z \right){E_a}\left( z \right)$, the expression above is further formulated as Eq. \eqref{E1} at the top of the next page.
\begin{figure*}
\begin{align}\label{E1}
  R_{est}^\infty  \approx& \frac{{\delta {m^m}}}{{2T\ln \left( 2 \right)\Gamma \left( m \right)}}\underbrace {\int_0^\infty  {\left( {1 + \frac{{md_t^{{\alpha _r}}}}{{{\Xi _{r,1}}x}}} \right){x^{m - 1}}\exp \left( { - mx} \right){E_1}\left( {\frac{{md_t^{{\alpha _r}}}}{{{\Xi _{r,1}}x}}} \right)dx} }_{{I_4}} \notag \\
 &+\sum\limits_{k = {\text{1}}}^{m - 1} {\frac{{\delta {m^m}}}{{2T\ln \left( 2 \right)\Gamma \left( m \right)}}} \underbrace {\int_0^\infty  {{x^{m - 1}}\exp \left( { - mx} \right)\mathop {{E_{k + 1}}}\limits_{{P_{BS}} \to \infty } \left( {\frac{{md_t^{{\alpha _r}}}}{{{\Xi _{r,1}}x}}} \right)dx} }_{{I_5}},
\end{align}
\hrulefill
\end{figure*}

Since we have $\mathop {{E_{k + 1}}}\limits_{{P_{BS}} \to \infty } \left( {\frac{{md_t^{{\alpha _r}}}}{{{\Xi _{r,1}}x}}} \right) \approx \frac{{{{\left( { - \frac{{md_t^{{\alpha _r}}}}{{{\Xi _{r,1}}x}}} \right)}^k}}}{{k!}}\left( {\psi \left( {k + 1} \right) - \ln \left( {\frac{{md_t^{{\alpha _r}}}}{{{\Xi _{r,1}}x}}} \right)} \right) - \sum\limits_{q = 0\& q \ne k}^{m - 1} {\frac{{{{\left( { - \frac{{md_t^{{\alpha _r}}}}{{{\Xi _{r,1}}x}}} \right)}^q}}}{{q!\left( {q - k} \right)}}} $ for $k>0$. The equation $I_4$ is further formulated as:
\begin{align}
  &{I_4} = \underbrace {\int_0^\infty  {{x^{m - 1}}\exp \left( { - mx} \right){E_1}\left( {\frac{{md_t^{{\alpha _r}}}}{{{\Xi _{r,1}}x}}} \right)dx} }_{{I_6}}\notag\\
 & \hspace*{0.1cm} + \underbrace {\frac{{md_t^{{\alpha _r}}}}{{{\Xi _{r,1}}}}\int_0^\infty  {{x^{m - 2}}\exp \left( { - mx} \right){E_1}\left( {\frac{{md_t^{{\alpha _r}}}}{{{\Xi _{r,1}}x}}} \right)dx} }_{{I_{7}}}  .
\end{align}

Based on the asymptotic expressions, respectively expressed as $\gamma \left( {m,t} \right) = \left( {m - 1} \right)! - \exp \left( { - t} \right)\sum\limits_{k = 0}^{m - 1} {\frac{{\left( {m - 1} \right)!}}{{k!}}{t^k}} $, ${E_n}\left( z \right) \approx \frac{{{{\left( { - z} \right)}^{n - 1}}}}{{\left( {n - 1} \right)!}}\left( {\psi \left( n \right)} \right. -$ $\left. {  \ln \left( z \right)} \right) - \sum\limits_{k = 0{\text{\& }}k \ne n - 1} {\frac{{{{\left( { - z} \right)}^k}}}{{k!\left( {1 - n + k} \right)}}} $ for $n>1$, and ${E_1}\left( z \right) \approx  - {C_\gamma } - \ln \left( z \right) + z$, ${I_6}$ and $I_{7}$ are derived as:
\begin{align}
  &{I_6} = \frac{{d_t^{{\alpha _r}}\Gamma \left( {m - 1} \right)}}{{{\Xi _{r,1}}{m^{m - 2}}}} \notag\\
  & \hspace*{0.1cm} - \frac{{\Gamma \left( m \right)}}{{{m^m}}}\left( {\log \left( {\frac{{{m^2}d_t^{{\alpha _r}}}}{{{\Xi _{r,1}}}}} \right) - {\psi ^{\left( 0 \right)}}\left( m \right) - {C_\gamma }} \right) ,\\
  &{I_{7}} = {\left( {\frac{{md_t^{{\alpha _r}}}}{{{\Xi _{r,1}}}}} \right)^2}\frac{{\Gamma \left( {m - 2} \right)}}{{{m^{m - 2}}}}- \frac{{md_t^{{\alpha _r}}}}{{{\Xi _{r,1}}}}\frac{{\Gamma \left( {m - 1} \right)}}{{{m^{m - 1}}}} \notag\\
  & \hspace*{0.1cm} \times \left( {\log \left( {\frac{{{m^2}d_t^{{\alpha _r}}}}{{{\Xi _{r,1}}}}} \right) - {\psi ^{\left( 0 \right)}}\left( {m - 1} \right) - {C_\gamma }} \right) .
\end{align}

Then, we can derive $I_5$ of Eq. \eqref{E1} by substituting the asymptotic expressions of ${\mathop {{E_{k + 1}}}\limits_{{P_{BS}} \to \infty } \left( {\frac{{md_t^{{\alpha _r}}}}{{{\Xi _{r,1}}x}}} \right)}$, Finally, we can substitute $I_4$ and $I_5$ into \eqref{E1} to obtain the final answer as Eq. \eqref{Cfinal}.

\vspace{3cm}

\begin{@twocolumnfalse}

\section*{\centering{\large{Further Proofs in the Paper Titled by ``Semi-Integrated-Sensing-and-Communication (Semi-ISaC): From OMA to NOMA"}}}

\end{@twocolumnfalse}

\vspace{0.5cm}
\begin{abstract}
  The new concept of semi-integrated-sensing-and-communication (Semi-ISaC) is proposed for next-generation cellular networks. Compared to the state-of-the-art, where the total bandwidth is used for integrated sensing and communication (ISaC), the proposed Semi-ISaC framework provides more freedom as it allows that a portion of the bandwidth is exclusively used for either wireless communication or radar detection, while the rest is for ISaC transmission. To enhance the bandwidth efficiency (BE), we investigate the evolution of Semi-ISaC networks from orthogonal multiple access (OMA) to non-orthogonal multiple access (NOMA). This paper provides the proofs of the journal version submitted to IEEE Transactions of Communications, namely ``Semi-Integrated-Sensing-and-Communication (Semi-ISaC): From OMA to NOMA". The proofs include Lemma 1-2, Theorem 1-2, and Corollary 1-2. The other lemmas, theorems, and corollaries have been comprehensively proved in the journal version or similar to the provided proofs.
\end{abstract}

\begin{IEEEkeywords}
Semi-integrated-sensing-and-communication, non-orthogonal multiple access, orthogonal multiple access, outage probability, ergodic radar estimation information rate
\end{IEEEkeywords}

\setcounter{section}{0}

\section{Proof~A: the Proof of Lemma~1}
\renewcommand{\theequation}{A.\arabic{equation}}
\setcounter{equation}{0}

We formulate the interference of the radar echo as ${I_R} = {P_{BS}}{\mathcal{P}_r}\left( {{d_r}} \right)$ $\times{\left| {{h_{r,d}}} \right|^2}{\left| {{h_{r,u}}} \right|^2}{\gamma ^2}{B^2}\sigma _\tau ^2$. In this expression, we have the small-scale fading parameters as two variables, namely $\left| {{h_{r,d}}} \right|^2$ and ${\left| {{h_{r,u}}} \right|^2}$. We note that both variables obey the Gamma distribution, whose probability density function (PDF) and cumulative distribution function (CDF) are expressed as
\begin{align}\label{pdf}
{f_{{{\left| {{h_{r,u}}} \right|}^2}}}\left( x \right)&={f_{{{\left| {{h_{r,d}}} \right|}^2}}}\left( x \right) = \frac{{{m^m}}}{{\Gamma \left( m \right)}}{x^{m - 1}}\exp \left( { - mx} \right),\\
{F_{{{\left| {{h_{r,u}}} \right|}^2}}}\left( x \right) &= {F_{{{\left| {{h_{r,d}}} \right|}^2}}}\left( x \right) = \frac{1}{{\Gamma \left( m \right)}}\gamma \left( {m,mx} \right),
\end{align}
where $\Gamma\left(\cdot\right)$ is the Gamma function and $\gamma\left(\cdot,\cdot\right)$ is the lower incomplete Gamma function.

Hence, the expectation of ${I_R}$ with respect to the variable $d_r$, namely $\mathbb{E}\left[ {{I_R}} \right]\left( {{d_r}} \right)$, is formulated as
\begin{align}
  \mathbb{E}\left[ {{I_R}} \right]\left( {{d_r}} \right) &= {P_{BS}}{\mathcal{P}_r}\left( {{d_r}} \right){\gamma ^2}{B^2}\sigma _\tau ^2\int_0^\infty  {x{f_{{{\left| {{h_{r,d}}} \right|}^2}}}\left( x \right)dx} \notag\\
  &\hspace*{0.1cm}\times \int_0^\infty  {y{f_{{{\left| {{h_{r,u}}} \right|}^2}}}\left( y \right)dy}  \notag \\
  & = {P_{BS}}{\mathcal{P}_r}\left( {{d_r}} \right){\gamma ^2}\beta _{semi}^2{B^2}\sigma _\tau ^2{\left( {\frac{{{m^m}}}{{\Gamma \left( m \right)}}} \right)^2} \notag\\
  &\hspace*{0.1cm}\times \int_0^\infty  {{x^m}\exp \left( { - mx} \right)dx} \int_0^\infty  {{y^m}\exp \left( { - my} \right)} dy, \notag \\
   &= {P_{BS}}{\mathcal{P}_r}\left( {{d_r}} \right){\gamma ^2}\beta _{semi}^2{B^2}\sigma _\tau ^2.
\end{align}

Finally, the proof of Lemma 1 is completed.

\section{Proof~B: the Proof of Theorem~1}
\renewcommand{\theequation}{B.\arabic{equation}}
\setcounter{equation}{0}

We have the signal-to-noise ratio (SNR) expression of the OMA user denoted as $\gamma _j^{OMA} = \frac{{{P_j}{\mathcal{P}_c}\left( {{d_j}} \right){{\left| {{h_j}} \right|}^2}}}{{{P_{BS}}{\mathcal{P}_r}\left( {{d_r}} \right){{\left| {{h_{r,d}}} \right|}^2}{{\left| {{h_{r,u}}} \right|}^2}{\gamma ^2}{B^2}\sigma _\tau ^2 + {\sigma ^2}}}$. Upon introducing the subscript of $j \in \{c,r\}$ for representing the communication transmitter and the radar target, the expressions of the outage probability (OP) and of the ergodic rate are respectively expressed as
\begin{align}
\mathbb{P}_j^{OMA}& = \Pr \left\{ {\gamma _j^{OMA} < \gamma _{th}^{OMA}} \right\},\\
R_j^{OMA} &= \frac{1}{2}{\log _2}\left( {1 + \gamma _j^{OMA}} \right),
\end{align}
where $\gamma _{th}^{OMA}$ is the threshold.

In the following, we will first derive the closed-form OP expression. Upon defining the parameter $\Omega  = $ $\frac{{m\left( {{P_{BS}}{\mathcal{P}_r}\left( {{d_r}} \right){\gamma ^2}\beta_{semi}^2{B^2}\sigma _\tau ^2 + {\sigma ^2}} \right)}}{{{P_j}{\mathcal{P}_c}\left( {{d_j}} \right)}}$, the OP is formulated as
\begin{align}
 & \mathbb{P}_j^{OMA} = \Pr \left\{ {\gamma _j^{OMA} < \gamma _{th}^{OMA}} \right\} \notag \\
   &\hspace*{0.1cm}= \Pr \left\{ {\frac{{{P_j}{\mathcal{P}_c}\left( {{d_j}} \right){{\left| {{h_j}} \right|}^2}}}{{{P_{BS}}{\mathcal{P}_r}\left( {{d_r}} \right){{\left| {{h_{r,d}}} \right|}^2}{{\left| {{h_{r,u}}} \right|}^2}{\gamma ^2}{B^2}\sigma _\tau ^2 + {\sigma ^2}}} < \gamma _{th}^{OMA}} \right\} \notag \\
   &\hspace*{0.1cm}= \Pr \left\{ {{{\left| {{h_j}} \right|}^2} < \frac{{\left( {{P_{BS}}{\mathcal{P}_r}\left( {{d_r}} \right){\gamma ^2}{B^2}\sigma _\tau ^2 + {\sigma ^2}} \right)\gamma _{th}^{OMA}}}{{{P_j}{\mathcal{P}_c}\left( {{d_j}} \right)}}} \right\} \notag \\
   &\hspace*{0.1cm}= \frac{{\gamma \left( {m,\frac{{m\left( {{P_{BS}}{\mathcal{P}_r}\left( {{d_r}} \right){\gamma ^2}{B^2}\sigma _\tau ^2 + {\sigma ^2}} \right)\gamma _{th}^{OMA}}}{{{P_j}{\mathcal{P}_c}\left( {{d_j}} \right)}}} \right)}}{{\Gamma \left( m \right)}} \notag \\
   &\hspace*{0.1cm}= \frac{{\gamma \left( {m,\Omega \gamma _{th}^{OMA}} \right)}}{{\Gamma \left( m \right)}}.
\end{align}

We then derive the closed-form expression of the ergodic rate, yielding the following expression:
\begin{align}
  R_j^{OMA} &= \frac{1}{2}{\log _2}\left( {1 + \gamma _j^{OMA}} \right) \notag \\
   &= \frac{1}{{2\ln 2}}\int_0^\infty  {\frac{{1 - \mathbb{P}_j^{OMA}\left( {\gamma _{th}^{OMA}} \right)}}{{1 + x}}dx}  \notag \\
 & \mathop  = \limits^{\left( a \right)} \frac{1}{{2\ln 2}}\sum\limits_{k = 0}^{m - 1} {\frac{{{\Omega ^k}}}{{k!}}} \int_0^\infty  {\frac{{{x^k}\exp \left( { - \Omega x} \right)}}{{1 + x}}dx}  \notag \\
 & \mathop  = \limits^{\left( b \right)} \frac{1}{{2\ln 2}}\sum\limits_{k = 0}^{m - 1} {\frac{{{\Omega ^k}}}{{k!}}} \exp \left( \Omega  \right)\Gamma \left( {k + 1} \right)\Gamma \left( { - k,\Omega } \right),
\end{align}
where we derive the step $(a)$ by exploiting the series expansion of the incomplete Gamma function denoted as $\gamma \left( {m,t} \right) = \left( {m - 1} \right)! - \exp \left( { - t} \right)\sum\limits_{k = 0}^{m - 1} {\frac{{\left( {m - 1} \right)!}}{{k!}}{t^k}} $. We then derive the step $(b)$ by exploiting the integral denoted as $\int_0^\infty  {\frac{{{x^a}}}{{1 + x}}} \exp \left( { - bx} \right) = \exp \left( b \right)\Gamma \left( {a + 1} \right)\Gamma \left( { - a,b} \right)$.

Finally, based on the relationship between the generalized exponential integral (namely ${E_n}\left( \cdot \right)$) and the upper incomplete Gamma function with negative parameters (denoted as $\Gamma \left( { - k,\Omega } \right) = \frac{{{E_{1 + k}}\left( \Omega  \right)}}{{{\Omega ^k}}}$), we formulate the final expression of the ergodic rate:
\begin{align}
R_j^{OMA} = \frac{1}{{2\ln 2}}\sum\limits_{k = 0}^{m - 1} {\exp \left( \Omega  \right){E_{1 + k}}\left( \Omega  \right)} .
\end{align}

Hence, the proof of the Theorem 1 is completed.

\section{Proof~C: the Proof of Lemma~2}
\renewcommand{\theequation}{C.\arabic{equation}}
\setcounter{equation}{0}

Since we have both ${{\left| {{h_{r,d}}} \right|}^2}$ and ${{\left| {{h_{r,u}}} \right|}^2}$ obeying the Gamma distribution, the PDF and CDF of the multiplication of both variables, namely ${f_{{{\left| {{h_{r,d}}} \right|}^2}{{\left| {{h_{r,u}}} \right|}^2}}}\left( x \right)$ and ${F_{{{\left| {{h_{r,d}}} \right|}^2}{{\left| {{h_{r,u}}} \right|}^2}}}\left( x \right)$, are expressed as
\begin{align}\label{CC1}
  {f_{{{\left| {{h_{r,d}}} \right|}^2}{{\left| {{h_{r,u}}} \right|}^2}}}\left( z \right) &= \int_0^\infty  {\frac{1}{x}{f_{{{\left| {{h_{r,d}}} \right|}^2}}}} \left( x \right){f_{{{\left| {{h_{r,u}}} \right|}^2}}}\left( {\frac{z}{x}} \right)dx \notag\\
 & \mathop {=}\limits^{\left( a \right)} \frac{{{m^{2m}}{z^{m - 1}}}}{{{{\left( {\Gamma \left( m \right)} \right)}^2}}}\int_0^\infty  {\frac{1}{x}\exp \left( { {-} m\left( {x {+} \frac{z}{x}} \right)} \right)} dx \notag \\
 & \mathop  = \limits^{\left( b \right)} \frac{{2{m^{2m}}}}{{{{\left( {\Gamma \left( m \right)} \right)}^2}}}{z^{m - 1}}{K_0}\left( {2m\sqrt z } \right),
\end{align}
where ${K_0}\left(  \cdot  \right)$ is the modified Bessel function of the third kind. The step $(a)$ is obtained by substituting the PDF of the Gamma distribution (Eq. \eqref{pdf}). The step $(b)$ is formulated by exploiting Eq.[2.3.6.7] of \cite{table} in the journal version.

Since we have the transformation between the modified Bessel function and the Meijer-G function expressed as ${K_v}\left( x \right) = \frac{1}{2}{G{_0^2}{_2^0}\left( {\frac{{{x^2}}}{4}\left| {{_{\frac{v}{2}}^ \cdot}{ _{\frac{{ - v}}{2}}^ \cdot} } \right.} \right)}$, where ${G{_p^m}{_q^n}\left( {\cdot \left| {_{\left( {{b_q}} \right)}^{\left( {{a_p}} \right)}} \right.} \right)}$ is the Meijer-G function, we further formulate the PDF expression as
\begin{align}\label{CC2}
{f_{{{\left| {{h_{r,d}}} \right|}^2}{{\left| {{h_{r,u}}} \right|}^2}}}\left( z \right) = \frac{{{m^{2m}}}}{{{{\left( {\Gamma \left( m \right)} \right)}^2}}}{z^{m - 1}}G{_0^2}{_2^0}\left( {{m^2}z\left| {{_0^ \cdot}{ _0^ \cdot} } \right.} \right).
\end{align}

Based on the PDF expressed as Eq. \eqref{CC2}, we then derive the CDF of ${f_{{{\left| {{h_{r,d}}} \right|}^2}{{\left| {{h_{r,u}}} \right|}^2}}}\left( x \right)$ as
\begin{align}\label{CC3}
  {F_{{{\left| {{h_{r,d}}} \right|}^2}{{\left| {{h_{r,u}}} \right|}^2}}}\left( x \right) &= \frac{{2{m^{2m}}}}{{{{\left( {\Gamma \left( m \right)} \right)}^2}}}\int_0^x {{z^{m - 1}}{K_0}\left( {2m\sqrt z } \right)} dz \notag \\
  &\mathop  = \limits^{\left( a \right)} \frac{{{m^{2m}}}}{{{{\left( {\Gamma \left( m \right)} \right)}^2}}}\int_0^x {{z^{m - 1}}G{_0^2}{_2^0}\left( {{m^2}z\left| {{_0^ \cdot}{ _0^ \cdot} } \right.} \right)} dz \notag \\
  &\mathop  = \limits^{\left( b \right)} \frac{{{m^{2m}}}}{{{{\left( {\Gamma \left( m \right)} \right)}^2}}}{x^m}G{_1^2}{_3^1}\left( {{m^2}x\left| {{_{0,0}^{1 - m}}{_{ - m}^ \cdot }} \right.} \right) \notag\\
 & \mathop  = \limits^{\left( c \right)} \frac{{G{_1^2}{_3^1}\left( {{m^2}x\left| {_{m,m,0}^1} \right.} \right)}}{{{{\left( {\Gamma \left( m \right)} \right)}^2}}},
\end{align}
where the step $(a)$ is derived by exploiting ${K_v}\left( x \right) = \frac{1}{2}{G{_0^2}{_2^0}\left( {\frac{{{x^2}}}{4}\left| {{_{\frac{v}{2}}^ \cdot}{ _{\frac{{ - v}}{2}}^ \cdot} } \right.} \right)}$. The step $(b)$ is derived by exploiting the integral of $\int{_0^y} {{x^{\alpha  - 1}}G{_p^m}{_q^n}\left( {wx\left| {_{\left( {{b_q}} \right)}^{\left( {{a_p}} \right)}} \right.} \right)dx}  = {y^\alpha }G{_{p + 1}^m}{_{q + 1}^{n + 1}}\left( {wy\left| {_{{b_1}, \cdots ,{b_m}, \cdots ; - \alpha ,{b_{m + 1,}} \cdots ,{b_q}}^{{a_1}, \cdots ,{a_n}, \cdots ,1 - \alpha ;{a_{n,}} \cdots ,{a_p}}} \right.} \right)$. Additionally, the step $(c)$ is formulated by using ${z^p}G{_p^m}{_q^n}\left( {z\left| {_{\left( {{b_q}} \right)}^{\left( {{a_p}} \right)}} \right.} \right) = {z^p}G{_p^m}{_q^n}\left( {z\left| {_{\left( {{b_q}} \right) + p}^{\left( {{a_p}} \right) + p}} \right.} \right)$.

Finally, we complete the proof of Lemma 2 by the final expressions, namely Eq. \eqref{CC1} and Eq. \eqref{CC3}.

\section{Proof~D: the Proof of Theorem~2 and Corollary~1}
\renewcommand{\theequation}{D.\arabic{equation}}
\setcounter{equation}{0}

We provide the proofs of Theorem~2 and Corollary~1 jointly in this section. For the analytical results of the radar echoes, the expression of the ergodic REIR is formulated directly by exploiting Eq. \eqref{CC3}, yielding
\begin{align}\label{DD1}
  &{R_{est}} \leqslant \frac{\delta }{{2T}}{\log _2}\left( {1 + {\Xi _{r,1}}{{\left( {{d_r}} \right)}^{ - {\alpha _r}}}{{\left| {{h_{r,d}}} \right|}^2}{{\left| {{h_{r,u}}} \right|}^2}} \right) \notag \\
  &\hspace*{0.1cm} = \frac{\delta }{{2T\ln \left( 2 \right)}}\int_0^\infty  {\frac{1}{{z + 1}}\left( {1 - {F_{{{\left| {{h_{r,d}}} \right|}^2}{{\left| {{h_{r,u}}} \right|}^2}}}\left( {\frac{{d_t^{{\alpha _r}}}}{{{\Xi _{r,1}}}}z} \right)} \right)} dz \notag \\
  &\hspace*{0.1cm} = \frac{\delta }{{2T\ln \left( 2 \right)}}\int_0^\infty  {\frac{1}{{z + 1}}\left( {1 - \frac{{G{_1^2}{_3^1}\left( {\frac{{{m^2}d_t^{{\alpha _r}}}}{{{\Xi _{r,1}}}}z\left| {_{m,m,0}^1} \right.} \right)}}{{{{\left( {\Gamma \left( m \right)} \right)}^2}}}} \right)} dz ,
\end{align}
where we have ${\Xi _{r,1}} = \frac{{2T{P_{BS}}{\gamma ^2}{B^2}\sigma _\tau ^2{G_r}{C_r}{{\left( {{d_r}} \right)}^{ - {\alpha _r}}}}}{{{k_B}{T_{temp}}}}$.

Upon considering $m=1$ as a special case, we further derive the above equation Eq. \eqref{DD1} as
\begin{align}\label{DD2}
  &{R_{est}} \leqslant \frac{\delta }{{2T}}{\log _2}\left( {1 + {\Xi _{r,1}}{{\left( {{d_r}} \right)}^{ - {\alpha _r}}}{{\left| {{h_{r,d}}} \right|}^2}{{\left| {{h_{r,u}}} \right|}^2}} \right) \notag \\
  &\hspace*{0.1cm}\mathop  = \limits^{\left( a \right)} \frac{\delta }{{2T\ln \left( 2 \right)}}\int_0^\infty  \int_0^\infty  \frac{1}{{z + 1}}\notag\\
  &\hspace*{0.1cm}\hspace*{0.1cm}\times\left( {1 - {F_{{{\left| {{h_{r,d}}} \right|}^2}}}\left( {\frac{{d_t^{{\alpha _r}}}}{{{\Xi _{r,1}}x}}z} \right)} \right) {f_{{{\left| {{h_{r,u}}} \right|}^2}}}\left( x \right)dxdz  \notag \\
  &\hspace*{0.1cm}\mathop  = \limits^{\left( b \right)} \frac{\delta }{{2T\ln \left( 2 \right)}}\int_0^\infty  {\int_0^\infty  {\frac{{\exp \left( { - \frac{{d_t^{{\alpha _r}}z}}{{{\Xi _{r,1}}x}}} \right)}}{{z + 1}}} dz\exp \left( { - x} \right)dx} \notag \\
  &\hspace*{0.1cm}\mathop  = \limits^{\left( c \right)} \frac{{ - \delta }}{{2T\ln \left( 2 \right)}}\int_0^\infty  {\exp } \left( {\frac{{d_t^{{\alpha _r}}}}{{{\Xi _{r,1}}x}}} \right){\rm{Ei}}\left( { - \frac{{d_t^{{\alpha _r}}}}{{{\Xi _{r,1}}x}}} \right) \exp \left( { - x} \right) dx\notag \\
  &\hspace*{0.1cm}\mathop  = \limits^{\left( d \right)} \frac{{ - \delta }}{{2T\ln \left( 2 \right)}}\int_0^\infty  {{x^{ - 2}}} \exp \left( {\frac{{d_t^{{\alpha _r}}x}}{{{\Xi _{r,1}}}} - \frac{1}{x}} \right){\rm{Ei}}\left( { - \frac{{d_t^{{\alpha _r}}}}{{{\Xi _{r,1}}}}x} \right) dx\notag \\
  &\hspace*{0.1cm}\mathop  = \limits^{\left( e \right)} \frac{\delta }{{2T\ln \left( 2 \right)}}G{_1^3}{_3^1}\left( {d_t^{{\alpha _r}}\Xi _{r,1}^{ - 1}\left| {_{0,0,1}^0} \right.} \right) ,
\end{align}
where $\rm{Ei}\left(\cdot\right)$ is the exponential integral. We express the probability of both variables by integrals, yielding the step $(a)$. Additionally, we substitute the PDF of the Gamma distribution (Eq. \eqref{pdf}) into the step $(a)$ of Eq. \eqref{DD2}, yielding the step $(b)$. Then, we exploit the Eq.[2.3.4.4] of \cite{table} to obtain the step $(c)$. We further derive the step $(d)$ by replacing $\frac{1}{x}$ by $x$. Finally, we have the step $(e)$ by using the integral of $\int_0^\infty  {{x^{ - 2}}} \exp \left( {Ax - \frac{1}{x}} \right)\rm{Ei}\left( { - Ax} \right)dx = G{_1^3}{_3^1}\left( {A\left| {_{0,0,1}^0} \right.} \right)$. And the proofs of Theorem~2 and Corollary~1 are completed.

\section{Proof~E: the Proof of Corollary~2}
\renewcommand{\theequation}{E.\arabic{equation}}
\setcounter{equation}{0}

Based on Theorem 3, we have the outage probability expression of the communication transmitter as
\begin{align}\label{EE1}
  \mathbb{P}_c^I =& 1 - \exp \left( { - \frac{{m{\gamma _{th}}}}{{{P_c}}}\left( {{a_1} + {a_2}} \right)} \right)\sum\limits_{p = 0}^{m - 1} {\frac{{{m^r}\gamma _{_{th}}^p}}{{\left( {m - 1} \right)!p!}}}  \notag\\
  & \times \sum\limits_{r = 0}^p {C_p^r\frac{{{{\left( {{a_1} + {a_2}} \right)}^r}{{\left( {{P_r}{a_3}} \right)}^{p - r}}}}{{P_c^p}}} \Gamma \left( {m + p - r} \right)\notag\\
  &\times {\left( {\frac{{{\gamma _{th}}{a_3}{P_r}}}{{{P_c}}} + 1} \right)^{ - \left( {m + p - r} \right)}} ,
\end{align}
where we have ${a_1} {=} \frac{{{P_{BS}}{G_r}{C_r}{{\left( {{d_r}} \right)}^{ - {\alpha _r}}}{\gamma ^2}{\beta_{semi}^2}{B^2}\sigma _\tau ^2}}{{{G_c}{C_c}{{\left( {{d_c}} \right)}^{ - {\alpha _c}}}}}$, ${a_2} {=} \frac{{{\sigma ^2}}}{{{G_c}{C_c}{{\left( {{d_c}} \right)}^{ - {\alpha _c}}}}}$, ${a_3} {=}  \frac{{{{\left( {{d_r}} \right)}^{ - {\alpha _c}}}}}{{{{\left( {{d_c}} \right)}^{ - {\alpha _c}}}}}$, and ${C_n^m}=n!/(m!(n-m)!)$.

In Scenario-I, the ergodic rate of the communication transmitter in the NOMA-based Semi-ISAC scenario is formulated as
\begin{align}\label{EE2}
R_c^{er,I} = \frac{1}{{\ln 2}}\int_0^\infty  {\frac{{1 - \mathbb{P}_c^I\left( {{\gamma _{th}}} \right)}}{{1 + {\gamma _{th}}}}} d{\gamma _{th}}.
\end{align}

We substitute Eq. \eqref{EE1} into Eq. \eqref{EE2}, yielding the further expression of the ergodic rate
\begin{align}\label{EE2}
  R_c^{er,I} & = \frac{1}{{\ln 2}}\sum\limits_{p = 0}^{m - 1} {\frac{{{m^r}}}{{\left( {m - 1} \right)!p!}}} \sum\limits_{r = 0}^p {C_p^r\frac{{{{\left( {{a_1} + {a_2}} \right)}^r}{{\left( {{P_r}{a_3}} \right)}^{p - r}}}}{{P_c^p}}} \notag\\
 &\hspace*{0.3cm}\times \Gamma \left( {m + p - r} \right) \int_0^\infty  {\frac{{{x^p}}}{{1 + x}}} {\left( {\frac{{x{a_3}{P_r}}}{{{P_c}}} + 1} \right)^{ - \left( {m + p - r} \right)}}\notag\\\
 &\hspace*{0.3cm}\times \exp \left( { - \frac{{mx}}{{{P_c}}}\left( {{a_1} + {a_2}} \right)} \right)dx \notag \\
 & \mathop  = \limits^{\left( a \right)} \frac{1}{{\ln 2}}\sum\limits_{p = 0}^{m - 1} {\frac{{{m^r}}}{{\left( {m - 1} \right)!p!}}} \sum\limits_{r = 0}^p {C_p^r\frac{{{{\left( {{a_1} + {a_2}} \right)}^r}\Gamma \left( {m {+} p {-} r} \right)}}{{P_c^p}}} \notag\\
 &\hspace*{0.3cm}\times {{\left( {{P_r}{a_3}} \right)}^{p - r}}  \sum\limits_{k = 0}^\infty  {\binom{  m {+} p {-} r {+} k {-} 1}{  k }} {\left( { - \frac{{{a_3}{P_r}}}{{{P_c}}}} \right)^k}\notag\\
 &\hspace*{0.3cm}\times  \int_0^\infty  {\frac{{{x^{p + k}}}}{{1 + x}}} \exp \left( { - \frac{{m\left( {{a_1} + {a_2}} \right)}}{{{P_c}}}x} \right)dx \notag \\
  & \mathop  = \limits^{\left( b \right)} \frac{1}{{\ln 2}}\sum\limits_{p = 0}^{m - 1} {\sum\limits_{r = 0}^p {C_p^r\frac{{\Lambda _1^{r - \left( {1 + p + k} \right)}{{\left( {{P_r}{a_3}} \right)}^{p - r}}}}{{\left( {m - 1} \right)!p!P_c^{p - r}}}} } \Gamma \left( {m {+} p {-} r} \right)\notag\\
  &\hspace*{0.3cm}\times\sum\limits_{k = 0}^\infty  {\binom{ m + p - r + k - 1 }{ k }} {\left( { - \frac{{{a_3}{P_r}}}{{{P_c}}}} \right)^k}\notag\\
  &\hspace*{0.3cm}\times\exp \left( {{\Lambda _1}} \right)\Gamma \left( {p + k + 1} \right){E_{1 + p + k}}\left( {{\Lambda _1}} \right) ,
\end{align}
where we have ${\Lambda _1} = \frac{{m\left( {{a_1} + {a_2}} \right)}}{{{P_c}}}$. The step $(a)$ is formulated by exploiting the binomial expansion with negative exponents, denoted as ${\left( {1 + x} \right)^{ - n}} = \sum\limits_{k = 0}^\infty  {\binom{  n + k - 1 }{  k }} {\left( { - x} \right)^k}$. Since we have the expression of ${E_n}\left( x \right) = {x^n}\Gamma \left( {1 - n,x} \right)$, the step $(b)$ is formulated by exploiting the following derivation:
\begin{align}\label{EE3}
 & \int_0^\infty  {\frac{{{x^{p + k}}}}{{1 + x}}} \exp \left( { - \frac{{m\left( {{a_1} + {a_2}} \right)}}{{{P_c}}}x} \right)dx \notag \\
   &\hspace*{0.3cm}= \exp \left( {\frac{{m\left( {{a_1} + {a_2}} \right)}}{{{P_c}}}} \right)\Gamma \left( {p + k + 1} \right)\notag\\
   &\hspace*{0.4cm} \times \Gamma \left( { - p - k,\frac{{m\left( {{a_1} + {a_2}} \right)}}{{{P_c}}}} \right) \notag \\
  &\hspace*{0.3cm} = \exp \left( {\frac{{m\left( {{a_1} + {a_2}} \right)}}{{{P_c}}}} \right)\Gamma \left( {p + k + 1} \right)\notag\\
  &\hspace*{0.4cm}\times {\left( {\frac{{m\left( {{a_1} + {a_2}} \right)}}{{{P_c}}}} \right)^{ - \left( {1 + p + k} \right)}}{E_{1 + p + k}}\left( {\frac{{m\left( {{a_1} + {a_2}} \right)}}{{{P_c}}}} \right) .
\end{align}

Hence, the proof of Corollary~2 is completed.

\bibliographystyle{IEEEtran}
\bibliography{mybib}

\end{document}